\documentclass[aps,pra,preprint,superscriptaddress,longbibliography]{revtex4-1}

\usepackage{amsmath}
\usepackage{amsthm}
\usepackage{amsfonts}
\usepackage{amssymb}
\usepackage{xcolor,graphicx}
\usepackage{tikz}
\usepackage{color}
\usepackage[pdftex]{hyperref} 
\usepackage{longtable}
\usepackage{array}
\usepackage{dsfont}

\begin{document}
	
	\title{Optical simulation of atomic decay enhancement and suppression}

	\author{B. Jaramillo-\'Avila}
	\email[e-mail: ]{jaramillo@inaoep.mx}
	\affiliation{CONACYT - Instituto Nacional de Astrof\'{i}sica, \'{O}ptica y Electr\'{o}nica, Calle Luis Enrique Erro No. 1. Sta. Ma. Tonantzintla, Pue. C.P. 72840, Mexico}
	
	\author{F. H. Maldonado-Villamizar}
	\email[e-mail: ]{fmaldonado@inaoep.mx}
	\affiliation{CONACYT - Instituto Nacional de Astrof\'{i}sica, \'{O}ptica y Electr\'{o}nica, Calle Luis Enrique Erro No. 1. Sta. Ma. Tonantzintla, Pue. C.P. 72840, Mexico}

	\author{B. M. Rodr\'iguez-Lara}
	\email[e-mail: ]{bmlara@tec.mx}
	\affiliation{Tecnologico de Monterrey, Escuela de Ingenier\'ia y Ciencias, Ave. Eugenio Garza Sada 2501, Monterrey, N.L., Mexico, 64849 }	
	
	\date{\today}
	
	\begin{abstract}
	We discuss the decay of a two-level system into an engineered reservoir of coupled harmonic oscillators in the single-excitation manifold and propose its optical simulation with an homogeneous chain of coupled waveguides where individual elements couple to an external waveguide.
	We use two approaches to study the decay of the optical analogue for the probability amplitude of the two-level system being in the excited state. 
	A Born approximation allows us to provide analytic closed-form amplitudes valid for small propagation distances.
	A Fourier-Laplace approach allows us to estimate an effective decay rate valid for long propagation distances.
	In general, our two analytic approximations match our numerical simulations using coupled mode theory and show non-Markovian decay into the engineered reservoir.
	In particular, we focus on two examples that provide enhancement or suppression of the decay decay rate using flat-top or Gaussian coupling distributions. 
	\end{abstract}
	
	
	\maketitle
\section{Introduction}

Engineered periodic photonic structures provide a robust and highly controllable platform to emulate a wide variety of quantum phenomena related to matter-radiation interactions using classical light \cite{Longhi2009, Rodriguez2018}. 
For example, there exist proposals for photonic analogies to Bloch oscillations \cite{Peschel1998, Morandotti1999, Pertsch1999, Lenz1999, Rodriguez2011, Villanueva2015}, quantum collapses and revivals \cite{Berry2001, Longhi2008}, atom-strong-field interactions \cite{Longhi2003, Marangoni2005, Longhi2005a}, Anderson localization \cite{Schwartz2008, Lahini2008, Thompson2010}, and various models of the Jaynes-Cummings type \cite{Longhi2011a, Crespi2012, RodriguezLara2014} among others. Such proposals use classical light propagating through arrays of 
waveguides described by coupled mode theory \cite{Snyder1972, McIntyre1973, Huang1994} and are amenable to experimental realization via laser inscription techniques \cite{Davis1996, Blomer2006, Szameit2010}.
These optical structures offer an immediate and accessible platform to study and visualize new characteristics of their quantum counterparts.

The decay of a quantum emitter coupled to a continuum is an interesting scenario concerning the interaction of matter and radiation. 
The spin-boson model \cite{Leggett1987, Weiss2009} is a well-known example of this,
\begin{align}
	\hat{H} = &
	\int_{0}^{\infty} d\omega~  
	\left\{ 
	\omega \hat{a}^{\dagger}(\omega) \hat{a}(\omega)
	+   g(\omega) \left[  \hat{\sigma}_{+} \hat{a}(\omega) + \hat{\sigma}_{-} \hat{a}^{\dagger}(\omega) \right] 
	\right\}
	+\frac{1}{2} \omega_{0} \hat{\sigma}_{z} .
\end{align}
It models a single two-level system, described by Pauli matrices $\hat{\sigma}_{z}$ and $\hat{\sigma}_{\pm}$ and frequency $\omega_{0}$, linearly coupled with strength $g(w)$ to an environment composed of a continuum of harmonic oscillators, described by creation (annihilation) operators $\hat{a}^{\dagger}$ ($\hat{a}$) and frequency $\omega$. 
This is the open quantum systems workhorse to study the effects of decoherence and non-Markovian dynamics \cite{Shiokawa2004, Guarnieri2016}.
The fact that it is possible to discretize and unfold the spin-boson model into that of a two-level system interacting with one end of a chain of coupled harmonic oscillators \cite{Vojta2005, Chin2010, Prior2010, Woods2014}, 
\begin{align}
	\hat{H} =& \frac{1}{2} \omega_{0} \hat{\sigma}_{z} + g \left(  \hat{\sigma}_{+} \hat{a}_{0} + \hat{\sigma}_{-} \hat{a}^{\dagger}_{0} \right)
	+  \sum_{j=0}^{\infty} \left[ \omega_{j} \hat{a}^{\dagger}_{j} \hat{a}_{j} + \gamma_{j}  \left( \hat{a}_{j}^{\dagger} \hat{a}_{j+1} + \hat{a}_{j+1} \hat{a}_{j}^{\dagger}  \right)   \right],
\end{align}
opens the door for the optical simulation of decay from an emitter into engineered environments using, for example, photonic lattices.
It allows the visualization and study of effects predicted to arise from impurities embedded within the geometric structure of atoms in a crystal \cite{Meade1995, Garanovich2012}; for example, the theoretical and experimental proposals to realize bound states \cite{Longhi2007, Plotnik2011}, decay control \cite{Dreisow2008, Longhi2009a}, or Zeno dynamics \cite{Longhi2006a, Longhi2007a, Biagioni2008}.

Here, we study the decay of an emitter into an engineered reservoir using an optical analogue for a two-level system coupled to the continuum of states given by a chain of identical oscillators \cite{Kockum2018, Longhi2020}. 
For the sake of simplicity, an external waveguide takes the role of the two-level emitter and we use the Bloch states of a chain of homogeneously coupled, identical waveguides as the optical analogue of the continuum.
We control the placement of some of the chain waveguides around the external waveguide to simulate the interaction between emitter and continuum with engineered coupling profiles leading to non-Markovian decay where we observe enhancement or suppression of the decay. 
In the following, we describe our quantum model and the continuum of states in the chain, Sec. \ref{Sec:2}. 
Then, we introduce our optical analogy using coupled mode theory and present two approaches to understand its dynamics, Sec. \ref{Sec:3}. 
One uses Born approximation to calculate short distance propagation of light in the system. 
The other uses Fourier-Laplace transform and yields an analytic expression for the leading effective decay rate. 
In Section \ref{Sec:4}, we compare our analytic predictions with numerical experiments to good agreement and demonstrate enhancement and suppression of the effective decay rate using two coupling distributions: flat-top and Gaussian. In addition, we show that this phenomenon is robust against noise. Finally, we summarize our findings and state our conclusions in Section \ref{Sec:5}.

\section{Quantum optics model}\label{Sec:2}

We focus on the analysis of a two-level system (TLS) interacting with coupled resonator optical waveguides (CROW)
\begin{align}
	\hat{H} =& \sum_{j=-\infty}^{\infty} \left[ \gamma  \left( \hat{a}_{j}^{\dagger} \hat{a}_{j+1} + \hat{a}_{j+1} \hat{a}_{j}^{\dagger}  \right)  +   g_{j} \left(  \hat{\sigma}_{+} \hat{a}_{j} + \hat{\sigma}_{-} \hat{a}^{\dagger}_{j} \right) \right]
	+ \delta \hat{\sigma}_{z},
\end{align}
where the resonators have identical frequency, $\omega$, and creation (annihilation) operators $\hat{a}_{j}^{\dagger}$ ($\hat{a}_{j}$).
We consider an homogeneous inter-resonator coupling strength $\gamma$ and a variable coupling strength between the $j$-th resonator and the TLS given by $g_{j}$.
This effective Hamiltonian rests in a frame defined by the total excitation number $\hat{N} = \sum_{j} \hat{a}_{j}^{\dagger} \hat{a}_{j} + \hat{\sigma}_{z}/2$ rotating at the frequency of the CROW resonators $\omega$, providing an effective detuning $\delta = (\omega_{0}-\omega)/2$.

As we are interested in the optical simulation of this quantum model, we study its dynamics in the single-excitation manifold, 
\begin{align}
	\vert \psi(t) \rangle = \mathcal{E}_{\alpha}(t) \vert e , 0 \rangle + \int_{-\pi}^{\pi} d\phi ~ \mathcal{E}_{\phi}(t) \vert g, \phi \rangle,
\end{align}
where $\vert e, 0 \rangle$ has the TLS in the excited state and the CROW in vacuum and $\vert g, \phi \rangle$ has the TLS in the ground state and the CROW in a single-excitation Bloch state,
\begin{align}
	\vert \phi \rangle = \frac{1}{\sqrt{2 \pi}} \sum_{k=-\infty}^{\infty} e^{i \phi k } \vert k \rangle,
\end{align}
where $\vert k \rangle$ has a single excitation in the $k$-th resonator and the rest in vacuum. 
Thus, we obtain equations of motion,
\begin{align}
	i \partial_{t} \mathcal{E}_{\alpha}(t) =&~ \delta \mathcal{E}_{\alpha}(t) + \int_{-\pi}^{\pi} d\varphi ~ G^{\ast}(\varphi) \mathcal{E}_{\varphi}(t), \\
	i \partial_{t} \mathcal{E}_{\phi}(t) =&~ \Omega(\phi) \mathcal{E}_{\phi}(t) + G(\phi) \mathcal{E}_{\alpha}(t),
\end{align}
for the probability amplitude of finding the excitation in the TLS, $\mathcal{E}_{\alpha}$, or in the CROW, $\mathcal{E}_{\phi}$. 
These amplitudes are given in terms of the effective dispersion relation for the Bloch modes and their coupling strength to the TLS,
\begin{align}
	\Omega(\phi) =&~ 2 \gamma \cos \phi, \\
	G(\phi) =&~ \frac{1}{\sqrt{2\pi}} \sum_{k=-\infty}^{\infty} g_{k} e^{i \phi k },
\end{align}
in that order.
It is straightforward to argue an optical analogy using classical fields in a whispering gallery mode CROW where an extra resonator is placed close to the CROW to act as the classical analogue of the TLS. 

\section{Coupled Mode Theory Model}\label{Sec:3}

Here, our interest lies on the optical simulation of the quantum optical system through arrays of evanescently coupled waveguides. 
The TLS is simulated by a single waveguide where $g_{j}$ denotes its coupling to the $j$-th waveguide in the chain.
The CROW is simulated by an infinite array of identical waveguides with homogeneous first-neighbors coupling $\gamma$. 
The detuning $\delta = \beta_{0} - \beta$ is the difference between the effective propagation constants of the external waveguide and those in the homogeneous chain.
A coupled mode theory analysis for these photonic lattices provides equivalent equations of motion, 
\begin{align}
	- i \partial_{z} \mathcal{E}_{\alpha}(z) =&~ \delta \mathcal{E}_{\alpha}(z) + \int_{-\pi}^{\pi} d\varphi ~ G^{\ast}(\varphi) \mathcal{E}_{\varphi}(z), \label{eq:AtomAnaloge}\\
	- i \partial_{z} \mathcal{E}_{\phi}(z) =&~ \Omega(\phi) \mathcal{E}_{\phi}(z) + G(\phi) \mathcal{E}_{\alpha}(z), \label{eq:FieldAnaloge}
\end{align}
for the modal field amplitudes in the external waveguide and the Bloch modes in the homogeneous chain, $\mathcal{E}_{\alpha}$ and $\mathcal{E}_{\phi}$.
These modal field amplitudes play the analogue role of probability amplitudes in the quantum model.
We favor photonic lattices as laser writing techniques allow for control of refractive index and placement position of individual waveguides in three dimensions \cite{Szameit2010,Gross2015}.
It may be possible to complicate an experimental realization to address, for example, engineered dispersion relations by treating inhomogeneous chains or complex coupling patterns that depend on the propagation direction.

Here, we are interested in simulating the decay of an atomic excitation into an engineered reservoir.
In order to provide an analytic guide, we follow an approach similar to that in the study of atomic decay into an oscillator reservoir. 
First, we take the equation for the optical analogue of the field probability amplitude and integrate it, 
\begin{align}
	\!\!\mathcal{E}_{\phi}(z) = e^{i \Omega(\phi) z} \mathcal{E}_{\phi}(0) +  i G(\phi) \!\! \int_{0}^{z} \!\!\! d\zeta \, e^{i \Omega(\phi) (z - \zeta)} \mathcal{E}_{\alpha}(\zeta).
\end{align}
Then, we substitute it into the equation for the optical analogue of the TLS excited state probability amplitude, 
\begin{align}
	\partial_{z} \mathcal{E}_{\alpha} (z) =&\, i \delta \mathcal{E}_{\alpha} (z) + i \int_{-\pi}^{\pi} d\varphi~ G^{\ast}(\varphi) e^{i \Omega(\varphi) z} \mathcal{E}_{\varphi}(0)
	- \int_{-\pi}^{\pi} d\varphi~ \int_{0}^{z}d\zeta~  \vert G(\varphi)\vert^2 e^{i \Omega(\varphi) (z-\zeta)} \mathcal{E}_{\alpha}(\zeta).
\end{align}
For the sake of simplicity, we focus on an initial condition set where the excitation starts at the waveguide playing the role of the TLS, 
\begin{align}\label{eq:InitialConditions}
	\mathcal{E}_{\alpha}(0) = 1 \qquad \mathrm{and} \qquad \mathcal{E}_{\phi}(0) = 0.
\end{align}
Under these conditions and upon substitution of all the involved parameters, we obtain an integro-differential equation, 
\begin{align}
	\partial_{z} \mathcal{E}_{\alpha} (z) =& - \!\!\!\! \sum_{j,k = -\infty}^{\infty} \!\! i^{\vert k-j \vert} g_{j}^{\ast} g_{k} \! \int_{0}^{z} \!\! d\zeta \, J_{\vert k-j \vert} \left[ 2 \gamma (z-\zeta) \right] \mathcal{E}_{\alpha}(\zeta)
	+ i \delta \mathcal{E}_{\alpha} (z),
\end{align}
in terms of a sum of Bessel functions of the first kind weighted by a product of the coupling strength between individual waveguides in the chain and the external waveguide.

The standard method to solve integro-differential equations of this type is, first, to solve for the analogue of the TLS state probability amplitude,
\begin{widetext}
	\begin{align}
		\mathcal{E}_{\alpha}(z)= e^{i \delta z} \left\{ 1 - \sum_{j,k = -\infty}^{\infty} i^{\vert k-j \vert} g_{j}^{\ast} g_{k}  \int_{0}^{z} d\zeta_{1}~ \int_{0}^{\zeta_{1}} d\zeta_{2}~ e^{-i \delta \zeta_{1}} J_{\vert k-j \vert} \left[ 2 \gamma (\zeta_{1}-\zeta_{2}) \right]  \mathcal{E}_{a}(\zeta_{2}) \right\}. 
	\end{align}
\end{widetext}
Then, iterate the integral term until the solution converges.
Here, we restrict ourselves to the scenario where the detuning in the individual propagation constants is larger than the coupling strength between waveguides in the chain and the external waveguide $\delta > g_{j}$.
This is an analogy to Born weak-coupling approximation in the quantum system. 
It yields an approximate solution,\\
\begin{widetext}
	\begin{align}
		\mathcal{E}_{\alpha}^{(1)}(z) \approx e^{i \delta z} \left\{ 1 - \sum_{j,k = -\infty}^{\infty} i^{\vert k-j \vert} g_{j}^{\ast} g_{k}  \int_{0}^{z} d\zeta_{1}~ \int_{0}^{\zeta_{1}} d\zeta_{2}~ e^{i \delta (\zeta_{2}-\zeta_{1})} J_{\vert k-j \vert} \left[ 2 \gamma (\zeta_{1}-\zeta_{2}) \right] \right\},
	\end{align}
\end{widetext}
where it is not possible to perform Markov approximation in the integral term as it is usually done in the standard atomic decay scenario.
Nevertheless, it is possible to solve the integral in the right hand side of this equation if we expand the exponential in its Maclaurin series. 
This result is valid for small propagation values and it is hard to extract some physical insight from its closed form.
Thus, we do not write it here. 
Instead, we discuss another approach to the solution that provides us with an approximation to the effective decay rate.

Let us start from the differential set in Eq.(\ref{eq:AtomAnaloge})-(\ref{eq:FieldAnaloge}) and perform a Fourier-Laplace transform \cite{Hormander},
\begin{align}\label{eq:FourierLaplace}
	\tilde{f}(\zeta)&=\int_{0}^{\infty}	e^{-i\zeta z} f(z) dz, 
\end{align}
that allows us to write the solution for the analogue of the TLS state probability amplitude under the atom decay conditions in Eq. (\ref{eq:InitialConditions}),
\begin{align}
	\tilde{\mathcal{E}}_{\alpha}(\zeta)&= \frac{-i}{\zeta-\delta-\Sigma(\zeta)},
\end{align}
where we use the shorthand notation for the coupling function
\begin{align}
	\Sigma(\zeta) 
	&= \int_{-\pi}^{\pi}\frac{\left|G(\phi)\right|^{2}}{\zeta-\Omega(\phi)}d\phi 
	\nonumber \\
	&=  -i \sum_{j,k=-\infty}^{\infty} \frac{ g_{j} g_{k}^{\ast}}{ \sqrt{ 4 \gamma^2 - \zeta^2 }}  e^{- i \vert j - k \vert \arccos \frac{\zeta}{2\gamma}}.
\end{align}
It is possible to calculate the inverse Fourier-Laplace transform,
\begin{align}
	f(z) = \frac{1}{2\pi}\int_{-\infty+i\epsilon}^{\infty+i\epsilon}e^{i\zeta z}\tilde{f}(\zeta)d\zeta,
\end{align}
using the formula \cite{Visuri2018},
\begin{align}
	\mathcal{E}_{\alpha}(z) =& -\frac{1}{\pi}\int_{-2\gamma}^{2\gamma} e^{i \zeta z} \Im \left[ \frac{1}{\zeta-\delta-\Sigma(\zeta)} \right] d\zeta 
	+ i 2 \sum_{k} \mathrm{Res}(\mathcal{\zeta}_{\alpha};z_{k}),
\end{align}
where the first term is an integral around the branch cut defined by the square root in the auxiliary function $\Sigma(\zeta)$ and the second is the sum of residues for the $\alpha$-th pole on the real line outside the branch cut.
As the detuning $\delta$ is real, the imaginary part of the coupling $\Sigma(\zeta)$ will rule the decay of the analogue of the TLS excited state probability amplitude and we can approximate,
\begin{eqnarray}
\mathcal{E}_{\alpha}(z) = e^{-\Gamma z} \mathcal{A}(z), \quad \mathrm{with} \quad \Gamma = \Im \left[ \Sigma(\delta) \right].
\end{eqnarray}
These results, the Born approximation for small propagation distances and the Fourier-Laplace method to approximate the effective decay, allow us to discuss particular examples where the decay can be enhanced or suppressed. 

\section{Enhancement and Suppression of Decay}\label{Sec:4}

\begin{figure*}
	\includegraphics{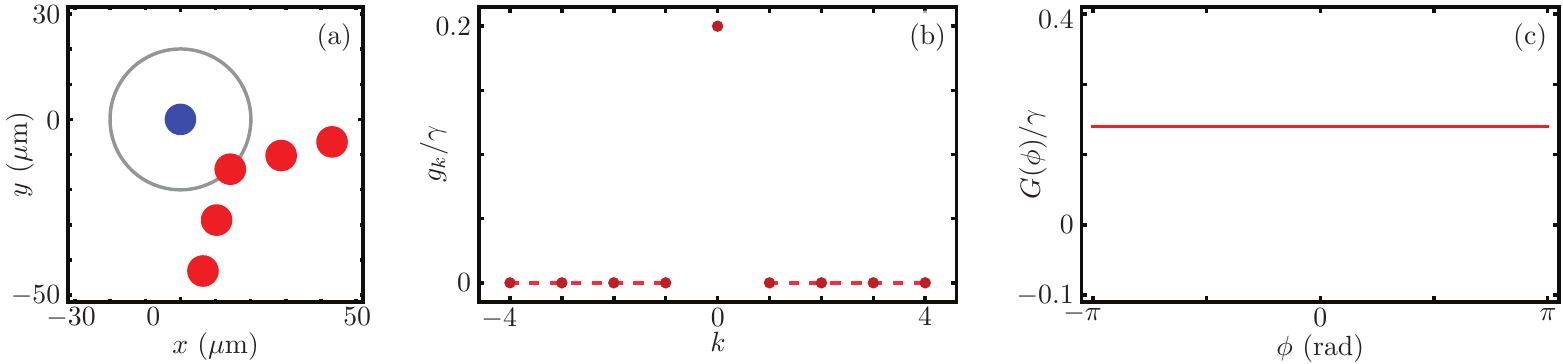}
	\caption{(a) Photonic lattice configuration, (b) coupling strength distribution between individual chain elements and the external waveguide in units of the chain coupling constant $g_{k}/\gamma$, and (c) effective coupling strength between the external waveguide to the Bloch modes of the homogeneous chain in units of the chain coupling constant $G(\phi)/\gamma$.} \label{fig:Fig1}
\end{figure*}

\begin{figure}
	\includegraphics{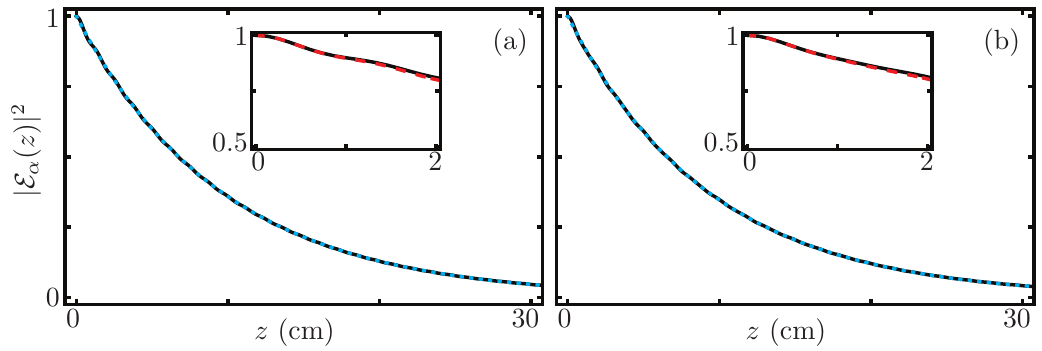}
	\caption{Optical analogue for the probability of finding the TLS in the excited state $\vert \mathcal{E}_{\alpha}(z)\vert^{2}$ (a) on-resonance $\delta=0$ and (b) off-resonance with $\delta = 0.5 \gamma$. We show coupled mode theory results (black solid line), numerical solution to the coupled integro-differential set (completely overlapped with black solid line), analytic solution using the Born approximation $\mathcal{E}_{\alpha}^{(3)}$ (red dashed line), analytic solution using the Fourier-Laplace approach (cyan dotted line) providing field amplitude decay rates (a) $\Gamma = 2 \times 10^{-2} ~\gamma$, (b) $\Gamma = 2.0656 \times 10^{-2} ~\gamma$.} \label{fig:Fig2}
\end{figure}

In order to provide practical examples, we consider a host of single-mode waveguides with circular profile in the weak-guiding regime. 
For the homogeneous chain, we use cores with refractive index $n_{\mathrm{co}}^{(\mathrm{ch})}=1.4479$ and radius $r_{\mathrm{co}}^{(\mathrm{ch})}=4.5~\mu\mathrm{m}$ embedded in cladding with refractive index $n_{\mathrm{cl}}=1.4440$.
Each core supports a single $\mathrm{LP}_{01}$ mode at wavelength $\lambda = 1550~\mathrm{nm}$, the telecommunications C-band.
For the waveguides in the homogeneous chain, we set the core to core separation at $d_{\mathrm{ch}} = 15~\mu\mathrm{m}$.
This yields a chain effective propagation constant and coupling strength $\beta = 5.85975\times 10^{6}~\mathrm{rad}/\mathrm{m}$ and $\gamma = 256.635~\mathrm{rad}/\mathrm{m}$, in that order.
Our numerical coupled mode theory simulations use an homogeneous chain of $501$ elements where the last $20$ waveguides at each end are lossy in order to suppress back-reflections due to finite size.

We focus on two types of coupling strength distributions, flat-top and Gaussian, between individual waveguides of the homogeneous chain and the external waveguide.
These are simple to realize in laser written photonic lattices.
For each coupling distribution, we study on- and off-resonant scenarios. 
In the former, the external waveguide is identical to those in the chain. 
In the latter, the external waveguide has refractive index $n_{\mathrm{co}}^{(\mathrm{e})} = 1.44794$ and radius $r_{\mathrm{co}}^{(\mathrm{e})}=4.5~\mu\mathrm{m}$ that yield an effective propagation constant $\beta_{0} = 5.85988 \times 10^{6}~\mathrm{rad}/\mathrm{m}$ and a detuning $\delta = 0.5 \gamma$. 

\begin{figure*}
	\includegraphics{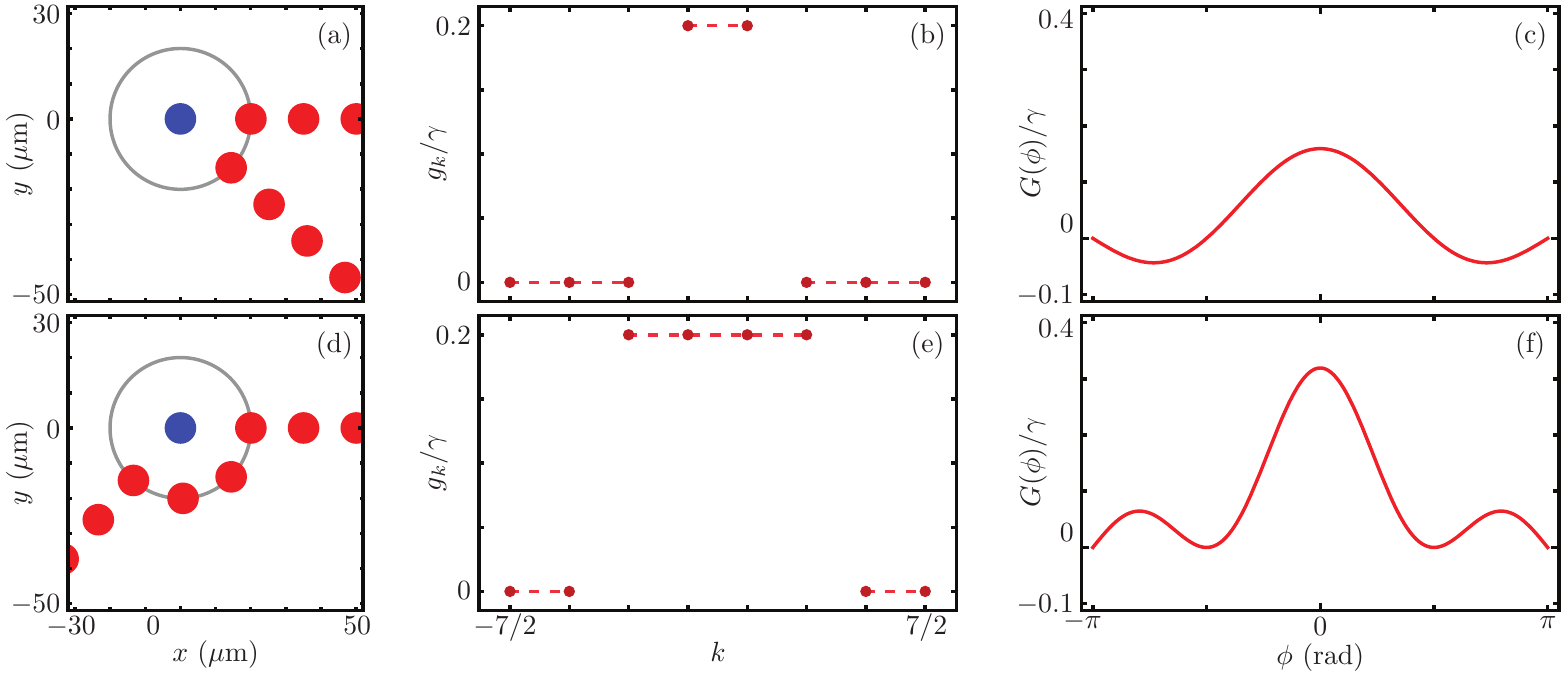}
	\caption{Same as Fig. \ref{fig:Fig1} for (a)-(c) two- and (d)-(f) four-element flat-top coupling distributions.} \label{fig:Fig3}
\end{figure*}

\begin{figure}
	\includegraphics{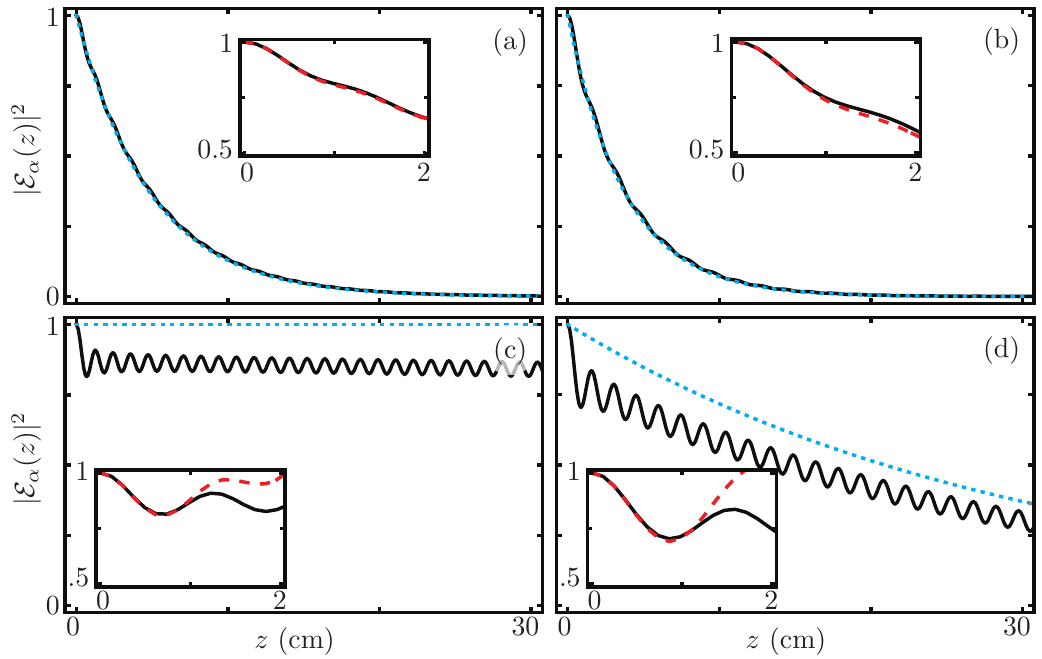}
	\caption{Optical analogue for the probability of finding the TLS in the excited state $\vert \mathcal{E}_{\alpha}(z)\vert^{2}$ for the two (a)-(b) and four element (c)-(d) flat-top coupling configurations with (a) and (c) on-resonance, $\delta=0$, and (b) and (d) off-resonance with $\delta = 0.5 \gamma$. We show coupled mode theory results (black solid line), numerical solution to the coupled integro-differential set (completely overlapped with black solid line), analytic solution using the Born approximation $\mathcal{E}_{\alpha}^{(3)}$ (red dashed line), analytic solution using the Fourier-Laplace approach (cyan dotted line) providing field amplitude decay rates (a) $\Gamma = 4 \times 10^{-2} ~\gamma$, (b) $\Gamma = 5.1640 \times 10^{-2} ~\gamma$, (c) $\Gamma = 0$ and (d) $\Gamma = 1.2910 \times 10^{-2} ~\gamma$.} \label{fig:Fig4}
\end{figure}

First, we consider a flat-top distribution for the coupling strengths,
\begin{align}
	g_{k} = g \sum_{p= q_{\mathrm{min}}}^{q_{\mathrm{max}}} \delta_{k,p}, 
\end{align}
where the homogeneous chain elements from position $q_{\mathrm{min}}$ to $q_{\mathrm{max}}$ form a circle of constant radius $r = 20.044~\mu\mathrm{m}$ around the external waveguide, Fig. \ref{fig:Fig1}(a), Fig. \ref{fig:Fig3}(a) and Fig. \ref{fig:Fig3}(d).
This yields a constant coupling strength $g = 0.2 \gamma$.
The scenario where just one waveguide from the homogeneous chain couples to the external waveguide belongs here, Fig. \ref{fig:Fig1}(a) and Fig. \ref{fig:Fig1}(b).
This provides an optical simulation of a TLS coupled to an engineered reservoir such that the coupling strength is constant for all continuous modes, Fig. \ref{fig:Fig1}(c).
The constant coupling strength $G(\phi)$ allows the use of Markov approximation in the resonant case to calculate the decay rate $\Gamma = g^{2}/ (2 \gamma) = 2 \times 10^{-2} ~\gamma$, Fig. \ref{fig:Fig2}(a), which is in accordance with that obtained in the Fourier-Laplace approach,
\begin{align}
	\Gamma(\delta) = \frac{g^{2}}{ \sqrt{4 \gamma^{2} - \delta^{2}} }.
\end{align} 
In this scenario, the off-resonant detuning in the propagation constants induces a slight increase in the effective field amplitude decay rate $\Gamma = 2.0656 \times 10^{-2} ~\gamma$, Fig. \ref{fig:Fig2}(b).

\begin{figure*}
	\includegraphics{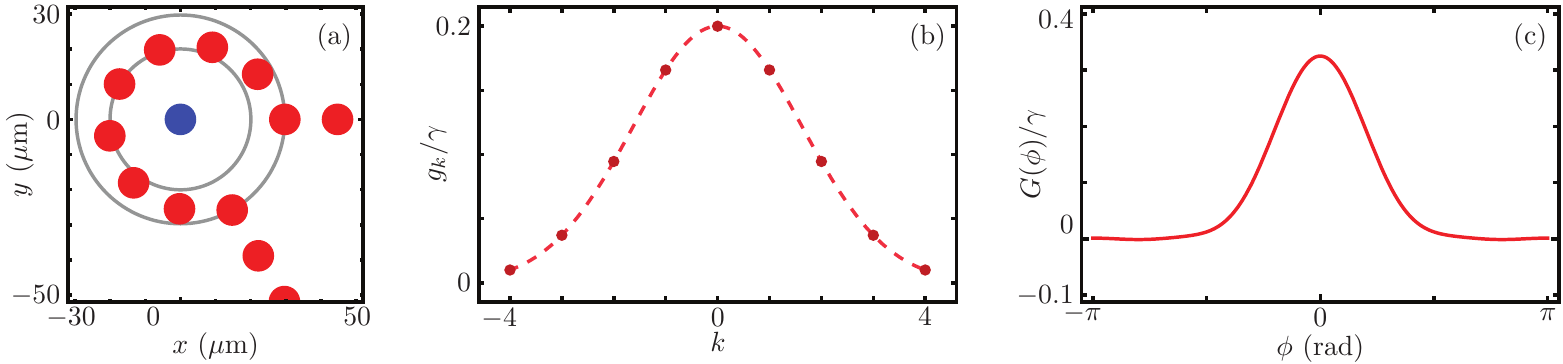}
	\caption{ Same as Fig. \ref{fig:Fig1} for Gaussian coupling distribution. } \label{fig:Fig5}
\end{figure*}

\begin{figure}
	\includegraphics{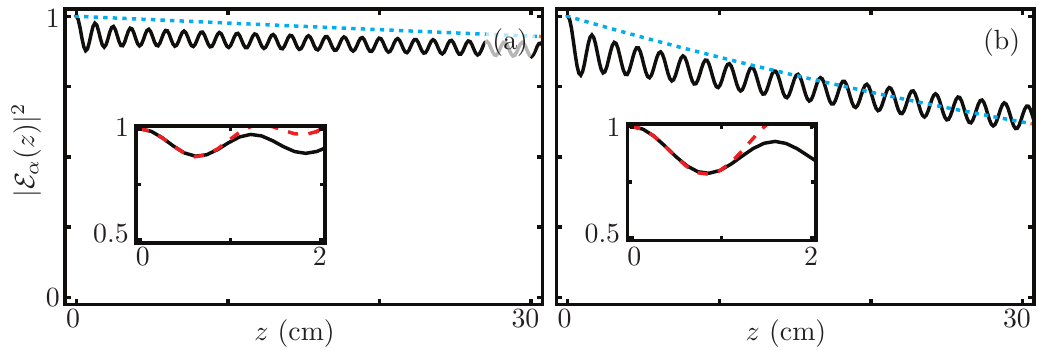}
	\caption{Same as Fig. \ref{fig:Fig3} for couplings given in Fig. \ref{fig:Fig4}. The field amplitude decay rates calculated with the Fourier-Laplace approach are (a) $\Gamma= 4.7591 \times 10^{-4} ~\gamma $ and (b) $\Gamma = 3.0660 \times 10^{-3} ~\gamma$. } \label{fig:Fig6}
\end{figure}

Coupling more elements from the homogeneous chain to the external waveguide, Fig. \ref{fig:Fig3}(a)-(b) and Fig. \ref{fig:Fig3}(d)-(e), simulates a reservoir whose effective coupling does not fulfill the requirements of Markov approximation, Fig. \ref{fig:Fig3}(c) and Fig. \ref{fig:Fig3}(f). 
The decay is no longer Markovian, still, it is possible to use Born approximation to good agreement in both on- and off-resonance scenarios, insets in Fig. \ref{fig:Fig4}.
For two coupled elements, we find an enhancement of the decay rate, compared to the single-waveguide coupling scenario, with the addition of a high frequency oscillation, $\Gamma = 4 \times 10^{-2} ~\gamma$ in Fig. \ref{fig:Fig4}(a) and $\Gamma = 5.1640 \times 10^{-2} ~\gamma$ in Fig. \ref{fig:Fig4}(b).
For four coupled elements, we find suppression of the decay rates, $\Gamma = 0$ in Fig. \ref{fig:Fig4}(c) and $\Gamma = 1.2910 \times 10^{-2} ~\gamma$ in Fig. \ref{fig:Fig4}(d).

Second, we implement a Gaussian distribution for the coupling strengths between elements of the homogeneous chain and the external waveguide,
\begin{align}
	g_{k} = g  e^{- \left(\frac{k -p_{0}}{\sigma} \right)^2},
\end{align}
where the distribution is centered at the $p_{0}$-th waveguide in the chain and its standard deviation is $\sigma$. 
For the sake of simplicity, we use the zeroth waveguide as the center of the distribution.
The homogeneous chain elements that couple to the external waveguide are distributed between the auxiliary circle defined above and a second auxiliary circle of radius $r =29.7087~\mu\mathrm{m}$, Fig. \ref{fig:Fig5}(a).
This provides us with couplings following a Gaussian distribution, Fig. \ref{fig:Fig5}(b), with standard deviation $\sigma=2.31105$ that simulates a reservoir whose effective coupling follows Fig. \ref{fig:Fig5}(c).
Again, the decay is non-Markovian, Fig. \ref{fig:Fig6}, and the decay rate is suppressed compared to the single-waveguide coupling case, $\Gamma= 4.7591 \times 10^{-4} ~\gamma$ in Fig. \ref{fig:Fig6}(a) and $\Gamma = 3.0660 \times 10^{-3} ~\gamma$ in Fig. \ref{fig:Fig6}(b). 

\begin{figure}
	\includegraphics{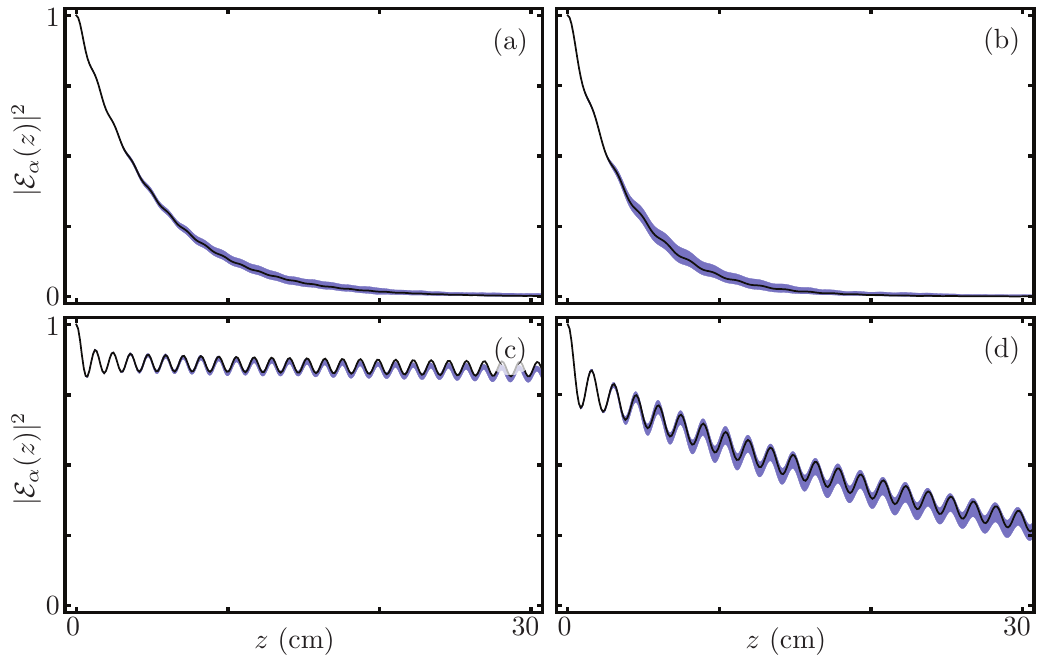}
	\caption{Same as Fig. \ref{fig:Fig4}, the black solid line shows the ideal evolution while the light blue region delimits one standard deviation above and below the average of 75 cases with up to $9\%$ random fluctuations in the coupling strength.}\label{fig:Fig7}
\end{figure}

Disorder in photonic structures leads to effects like Anderson localization of light \cite{DeRaedt1989,Schwartz2007,Segev2013} or crosstalk suppression \cite{Jaramillo2019}. 
Such disorder may arise from manufacturing circumstances; for example, the step precision in the motor controlling laser writing stages \cite{Alberucci2020}. 
For the sake of completeness, we study the effect of random fluctuations in our proposal. 
In particular, we add $z$-dependent random fluctuations to all the couplings in the chain using a spatial frequency of $81.6895~\mathrm{m}^{-1}$ and a maximum fluctuation amplitude of $9\%$, related to deviations of up to $250~\mathrm{nm}$ from the ideal position of the waveguides \cite{Alberucci2020}. 
Figure \ref{fig:Fig7} shows the ideal evolution from Fig. \ref{fig:Fig4} and compares it with the region delimited by one standard deviation above and below the average for the evolution of 75 independent cases with such random fluctuations. 
Decay enhancement or suppression remains even for individual realizations including this type of fabrication imprecision.

\begin{figure}
	\includegraphics{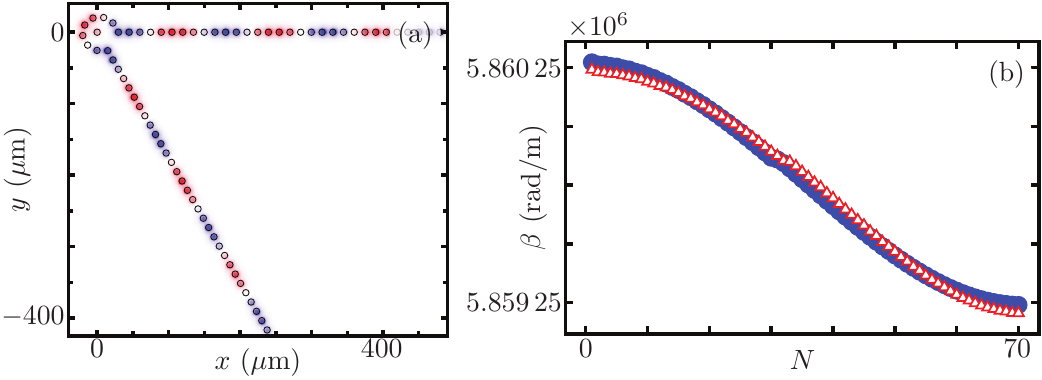}
	\caption{(a) FEM simulation of a normal mode and (b) comparison between effective mode propagation constants from CMT (filled blue dots) and FEM (empty red triangles) for the Gaussian profile in Fig. \ref{fig:Fig5}(a), with a total of 70 waveguides.} \label{fig:Fig8}
\end{figure}

In addition to the comparison between our analytic approximations and numerical results, we compared between the normal modes calculated using coupled mode theory and a finite element model (FEM) simulation of a smaller system, with only $70$ waveguides, see Fig. \ref{fig:Fig8}. We found good agreement between these two approaches further informing ourselves on the validity of our parameter values.

\section{Conclusion}\label{Sec:5}

We study the decay of a two-level system into an engineered continuum reservoir produced by an infinite chain of oscillators. 
In order to propose an optical analogy using a photonic lattice, we restrict ourselves to the single-excitation manifold.
The field amplitude in an external waveguide plays the role of the probability amplitude to find the two-level system in the excited state. 
The field in a chain of homogeneously coupled waveguides plays the role of the two-level system excitation decaying into an engineered continuum with nonlinear dispersion relation.
In addition, we control the placement of the waveguides in the homogeneous chain with respect to the external one to produce different coupling profiles that simulate different coupling profiles between the two-level 
system and the engineered continuum. 
Our optical simulation may be experimentally realized by arrays of coupled laser inscribed waveguides.

We explore different coupling profiles. 
In particular, we discuss flat-top and Gaussian distributions and find non-Markovian decay that leads to enhancement or suppression of the decay rate compared to standard Markovian decay. 
We implement two analytic approaches that allow us to calculate short-distance propagation and long-distance effective decay rate. 
We validate these approximations with coupled mode theory numerical simulations for parameters from telecomm C-band experiments.

Exploring the interaction of a quantum system with its environment is a demanding task as these systems are hard to realize in a controlled manner. 
Optical analogues that simulate reservoir and coupling engineering may aid in this exploration, as they are easier to implement and control in the laboratory, and can provide a platform to benchmark models for system-environment dynamics.

\begin{acknowledgments}
	B.J.-A. and F.H.M.-V. acknowledge support from CONACYT C\'atedra grupal No.~551. 
	All authors are profoundly indebted to Julio Abraham Mendoza-Fierro for valuable discussion and technical insights during the realization of this manuscript.
\end{acknowledgments}

\section*{References}

\begin{thebibliography}{52}%
	\makeatletter
	\providecommand \@ifxundefined [1]{%
		\@ifx{#1\undefined}
	}%
	\providecommand \@ifnum [1]{%
		\ifnum #1\expandafter \@firstoftwo
		\else \expandafter \@secondoftwo
		\fi
	}%
	\providecommand \@ifx [1]{%
		\ifx #1\expandafter \@firstoftwo
		\else \expandafter \@secondoftwo
		\fi
	}%
	\providecommand \natexlab [1]{#1}%
	\providecommand \enquote  [1]{``#1''}%
	\providecommand \bibnamefont  [1]{#1}%
	\providecommand \bibfnamefont [1]{#1}%
	\providecommand \citenamefont [1]{#1}%
	\providecommand \href@noop [0]{\@secondoftwo}%
	\providecommand \href [0]{\begingroup \@sanitize@url \@href}%
	\providecommand \@href[1]{\@@startlink{#1}\@@href}%
	\providecommand \@@href[1]{\endgroup#1\@@endlink}%
	\providecommand \@sanitize@url [0]{\catcode `\\12\catcode `\$12\catcode
		`\&12\catcode `\#12\catcode `\^12\catcode `\_12\catcode `\%12\relax}%
	\providecommand \@@startlink[1]{}%
	\providecommand \@@endlink[0]{}%
	\providecommand \url  [0]{\begingroup\@sanitize@url \@url }%
	\providecommand \@url [1]{\endgroup\@href {#1}{\urlprefix }}%
	\providecommand \urlprefix  [0]{URL }%
	\providecommand \Eprint [0]{\href }%
	\providecommand \doibase [0]{http://dx.doi.org/}%
	\providecommand \selectlanguage [0]{\@gobble}%
	\providecommand \bibinfo  [0]{\@secondoftwo}%
	\providecommand \bibfield  [0]{\@secondoftwo}%
	\providecommand \translation [1]{[#1]}%
	\providecommand \BibitemOpen [0]{}%
	\providecommand \bibitemStop [0]{}%
	\providecommand \bibitemNoStop [0]{.\EOS\space}%
	\providecommand \EOS [0]{\spacefactor3000\relax}%
	\providecommand \BibitemShut  [1]{\csname bibitem#1\endcsname}%
	\let\auto@bib@innerbib\@empty
	\bibitem [{\citenamefont {Longhi}(2009{\natexlab{a}})}]{Longhi2009}%
	\BibitemOpen
	\bibfield  {author} {\bibinfo {author} {\bibfnamefont {S.}~\bibnamefont
			{Longhi}},\ }\bibfield  {title} {\enquote {\bibinfo {title} {Quantum-optical
				analogies using photonic structures},}\ }\href {\doibase
		10.1002/lpor.200810055} {\bibfield  {journal} {\bibinfo  {journal} {Laser
				Photonics Rev.}\ }\textbf {\bibinfo {volume} {3}},\ \bibinfo {pages} {243}
		(\bibinfo {year} {2009}{\natexlab{a}})}\BibitemShut {NoStop}%
	\bibitem [{\citenamefont {Rodr\'iguez-Lara}\ \emph {et~al.}(2018)\citenamefont
		{Rodr\'iguez-Lara}, \citenamefont {El-Ganainy},\ and\ \citenamefont
		{Guerrero}}]{Rodriguez2018}%
	\BibitemOpen
	\bibfield  {author} {\bibinfo {author} {\bibfnamefont {B.~M.}\ \bibnamefont
			{Rodr\'iguez-Lara}}, \bibinfo {author} {\bibfnamefont {R.}~\bibnamefont
			{El-Ganainy}}, \ and\ \bibinfo {author} {\bibfnamefont {J.}~\bibnamefont
			{Guerrero}},\ }\bibfield  {title} {\enquote {\bibinfo {title} {Symmetry in
				optics and photonics: a group theory approach},}\ }\href {\doibase
		10.1016/j.scib.2017.12.020} {\bibfield  {journal} {\bibinfo  {journal} {Sci.
				Bull.}\ }\textbf {\bibinfo {volume} {63}},\ \bibinfo {pages} {244} (\bibinfo
		{year} {2018})},\ \Eprint {http://arxiv.org/abs/1803.00121} {arXiv:1803.00121
		[physics.optics]} \BibitemShut {NoStop}%
	\bibitem [{\citenamefont {Peschel}\ \emph {et~al.}(1998)\citenamefont
		{Peschel}, \citenamefont {Pertsch},\ and\ \citenamefont
		{Lederer}}]{Peschel1998}%
	\BibitemOpen
	\bibfield  {author} {\bibinfo {author} {\bibfnamefont {U.}~\bibnamefont
			{Peschel}}, \bibinfo {author} {\bibfnamefont {T.}~\bibnamefont {Pertsch}}, \
		and\ \bibinfo {author} {\bibfnamefont {F.}~\bibnamefont {Lederer}},\
	}\bibfield  {title} {\enquote {\bibinfo {title} {Optical {B}loch oscillations
				in waveguide arrays},}\ }\href {\doibase 10.1364/OL.23.001701} {\bibfield
		{journal} {\bibinfo  {journal} {Opt. Lett.}\ }\textbf {\bibinfo {volume}
			{23}},\ \bibinfo {pages} {1701} (\bibinfo {year} {1998})}\BibitemShut
	{NoStop}%
	\bibitem [{\citenamefont {Morandotti}\ \emph {et~al.}(1999)\citenamefont
		{Morandotti}, \citenamefont {Peschel}, \citenamefont {Aitchison},
		\citenamefont {Eisenberg},\ and\ \citenamefont
		{Silberberg}}]{Morandotti1999}%
	\BibitemOpen
	\bibfield  {author} {\bibinfo {author} {\bibfnamefont {R.}~\bibnamefont
			{Morandotti}}, \bibinfo {author} {\bibfnamefont {U.}~\bibnamefont {Peschel}},
		\bibinfo {author} {\bibfnamefont {J.~S.}\ \bibnamefont {Aitchison}}, \bibinfo
		{author} {\bibfnamefont {H.~S.}\ \bibnamefont {Eisenberg}}, \ and\ \bibinfo
		{author} {\bibfnamefont {Y.}~\bibnamefont {Silberberg}},\ }\bibfield  {title}
	{\enquote {\bibinfo {title} {Experimental observation of linear and nonlinear
				optical {B}loch oscillations},}\ }\href {\doibase
		10.1103/PhysRevLett.83.4756} {\bibfield  {journal} {\bibinfo  {journal}
			{Phys. Rev. Lett.}\ }\textbf {\bibinfo {volume} {83}},\ \bibinfo {pages}
		{4756} (\bibinfo {year} {1999})}\BibitemShut {NoStop}%
	\bibitem [{\citenamefont {Pertsch}\ \emph {et~al.}(1999)\citenamefont
		{Pertsch}, \citenamefont {Dannberg}, \citenamefont {Elflein}, \citenamefont
		{Br\"auer},\ and\ \citenamefont {Lederer}}]{Pertsch1999}%
	\BibitemOpen
	\bibfield  {author} {\bibinfo {author} {\bibfnamefont {T.}~\bibnamefont
			{Pertsch}}, \bibinfo {author} {\bibfnamefont {P.}~\bibnamefont {Dannberg}},
		\bibinfo {author} {\bibfnamefont {W.}~\bibnamefont {Elflein}}, \bibinfo
		{author} {\bibfnamefont {A.}~\bibnamefont {Br\"auer}}, \ and\ \bibinfo
		{author} {\bibfnamefont {F.}~\bibnamefont {Lederer}},\ }\bibfield  {title}
	{\enquote {\bibinfo {title} {Optical {B}loch oscillations in temperature
				tuned waveguide arrays},}\ }\href {\doibase 10.1103/PhysRevLett.83.4752}
	{\bibfield  {journal} {\bibinfo  {journal} {Phys. Rev. Lett.}\ }\textbf
		{\bibinfo {volume} {83}},\ \bibinfo {pages} {4752} (\bibinfo {year}
		{1999})}\BibitemShut {NoStop}%
	\bibitem [{\citenamefont {Lenz}\ \emph {et~al.}(1999)\citenamefont {Lenz},
		\citenamefont {Talanina},\ and\ \citenamefont {{de Sterke}}}]{Lenz1999}%
	\BibitemOpen
	\bibfield  {author} {\bibinfo {author} {\bibfnamefont {G.}~\bibnamefont
			{Lenz}}, \bibinfo {author} {\bibfnamefont {I.}~\bibnamefont {Talanina}}, \
		and\ \bibinfo {author} {\bibfnamefont {C.~M.}\ \bibnamefont {{de Sterke}}},\
	}\bibfield  {title} {\enquote {\bibinfo {title} {{B}loch oscillations in an
				array of curved optical waveguides},}\ }\href {\doibase
		10.1103/PhysRevLett.83.963} {\bibfield  {journal} {\bibinfo  {journal} {Phys.
				Rev. Lett.}\ }\textbf {\bibinfo {volume} {83}},\ \bibinfo {pages} {963}
		(\bibinfo {year} {1999})}\BibitemShut {NoStop}%
	\bibitem [{\citenamefont {Rodr\'iguez-Lara}(2011)}]{Rodriguez2011}%
	\BibitemOpen
	\bibfield  {author} {\bibinfo {author} {\bibfnamefont {B.~M.}\ \bibnamefont
			{Rodr\'iguez-Lara}},\ }\bibfield  {title} {\enquote {\bibinfo {title} {Exact
				dynamics of finite {G}lauber-{F}ock photonic lattices},}\ }\href {\doibase
		10.1103/PhysRevA.84.053845} {\bibfield  {journal} {\bibinfo  {journal} {Phys.
				Rev. A}\ }\textbf {\bibinfo {volume} {84}},\ \bibinfo {pages} {053845}
		(\bibinfo {year} {2011})},\ \Eprint {http://arxiv.org/abs/1108.3004}
	{arXiv:1108.3004 [quant-ph]} \BibitemShut {NoStop}%
	\bibitem [{\citenamefont {{Villanueva Vergara}}\ and\ \citenamefont
		{Rodr\'iguez-Lara}(2015)}]{Villanueva2015}%
	\BibitemOpen
	\bibfield  {author} {\bibinfo {author} {\bibfnamefont {L.}~\bibnamefont
			{{Villanueva Vergara}}}\ and\ \bibinfo {author} {\bibfnamefont {B.~M.}\
			\bibnamefont {Rodr\'iguez-Lara}},\ }\bibfield  {title} {\enquote {\bibinfo
			{title} {{G}ilmore-{P}erelomov symmetry based approach to photonic
				lattices},}\ }\href {\doibase 10.1364/OE.23.022836} {\bibfield  {journal}
		{\bibinfo  {journal} {Opt. Express}\ }\textbf {\bibinfo {volume} {23}},\
		\bibinfo {pages} {22836} (\bibinfo {year} {2015})},\ \Eprint
	{http://arxiv.org/abs/1506.02062} {arXiv:1506.02062 [physics.optics]}
	\BibitemShut {NoStop}%
	\bibitem [{\citenamefont {Berry}\ \emph {et~al.}(2001)\citenamefont {Berry},
		\citenamefont {Marzoli},\ and\ \citenamefont {Schleich}}]{Berry2001}%
	\BibitemOpen
	\bibfield  {author} {\bibinfo {author} {\bibfnamefont {M.}~\bibnamefont
			{Berry}}, \bibinfo {author} {\bibfnamefont {I.}~\bibnamefont {Marzoli}}, \
		and\ \bibinfo {author} {\bibfnamefont {W.}~\bibnamefont {Schleich}},\
	}\bibfield  {title} {\enquote {\bibinfo {title} {Quantum carpets, carpets of
				light},}\ }\href {\doibase 10.1088/2058-7058/14/6/30} {\bibfield  {journal}
		{\bibinfo  {journal} {Phys. World}\ }\textbf {\bibinfo {volume} {14}},\
		\bibinfo {pages} {39} (\bibinfo {year} {2001})}\BibitemShut {NoStop}%
	\bibitem [{\citenamefont {Longhi}(2008)}]{Longhi2008}%
	\BibitemOpen
	\bibfield  {author} {\bibinfo {author} {\bibfnamefont {S.}~\bibnamefont
			{Longhi}},\ }\bibfield  {title} {\enquote {\bibinfo {title} {Quantum bouncing
				ball on a lattice: An optical realization},}\ }\href {\doibase
		10.1103/PhysRevA.77.035802} {\bibfield  {journal} {\bibinfo  {journal} {Phys.
				Rev. A}\ }\textbf {\bibinfo {volume} {77}},\ \bibinfo {pages} {035802}
		(\bibinfo {year} {2008})}\BibitemShut {NoStop}%
	\bibitem [{\citenamefont {Longhi}\ \emph {et~al.}(2003)\citenamefont {Longhi},
		\citenamefont {Janner}, \citenamefont {Marano},\ and\ \citenamefont
		{Laporta}}]{Longhi2003}%
	\BibitemOpen
	\bibfield  {author} {\bibinfo {author} {\bibfnamefont {S.}~\bibnamefont
			{Longhi}}, \bibinfo {author} {\bibfnamefont {D.}~\bibnamefont {Janner}},
		\bibinfo {author} {\bibfnamefont {M.}~\bibnamefont {Marano}}, \ and\ \bibinfo
		{author} {\bibfnamefont {P.}~\bibnamefont {Laporta}},\ }\bibfield  {title}
	{\enquote {\bibinfo {title} {Quantum-mechanical analogy of beam propagation
				in waveguides with a bent axis: Dynamic-mode stabilization and radiation-loss
				suppression},}\ }\href {\doibase 10.1103/PhysRevE.67.036601} {\bibfield
		{journal} {\bibinfo  {journal} {Phys. Rev. E}\ }\textbf {\bibinfo {volume}
			{67}},\ \bibinfo {pages} {036601} (\bibinfo {year} {2003})}\BibitemShut
	{NoStop}%
	\bibitem [{\citenamefont {Marangoni}\ \emph {et~al.}(2005)\citenamefont
		{Marangoni}, \citenamefont {Janner}, \citenamefont {Ramponi}, \citenamefont
		{Laporta}, \citenamefont {Longhi}, \citenamefont {Cianci},\ and\
		\citenamefont {Foglietti}}]{Marangoni2005}%
	\BibitemOpen
	\bibfield  {author} {\bibinfo {author} {\bibfnamefont {M.}~\bibnamefont
			{Marangoni}}, \bibinfo {author} {\bibfnamefont {D.}~\bibnamefont {Janner}},
		\bibinfo {author} {\bibfnamefont {R.}~\bibnamefont {Ramponi}}, \bibinfo
		{author} {\bibfnamefont {P.}~\bibnamefont {Laporta}}, \bibinfo {author}
		{\bibfnamefont {S.}~\bibnamefont {Longhi}}, \bibinfo {author} {\bibfnamefont
			{E.}~\bibnamefont {Cianci}}, \ and\ \bibinfo {author} {\bibfnamefont
			{V.}~\bibnamefont {Foglietti}},\ }\bibfield  {title} {\enquote {\bibinfo
			{title} {Beam dynamics and wave packet splitting in a periodically curved
				optical waveguide: Multimode effects},}\ }\href {\doibase
		10.1103/PhysRevE.72.026609} {\bibfield  {journal} {\bibinfo  {journal} {Phys.
				Rev. E}\ }\textbf {\bibinfo {volume} {72}},\ \bibinfo {pages} {026609}
		(\bibinfo {year} {2005})}\BibitemShut {NoStop}%
	\bibitem [{\citenamefont {Longhi}(2005)}]{Longhi2005a}%
	\BibitemOpen
	\bibfield  {author} {\bibinfo {author} {\bibfnamefont {S.}~\bibnamefont
			{Longhi}},\ }\bibfield  {title} {\enquote {\bibinfo {title} {Wave packet
				dynamics in a helical optical waveguide},}\ }\href {\doibase
		10.1103/PhysRevA.71.055402} {\bibfield  {journal} {\bibinfo  {journal} {Phys.
				Rev. A}\ }\textbf {\bibinfo {volume} {71}},\ \bibinfo {pages} {055402}
		(\bibinfo {year} {2005})}\BibitemShut {NoStop}%
	\bibitem [{\citenamefont {Schwartz}\ \emph {et~al.}(2008)\citenamefont
		{Schwartz}, \citenamefont {Fishman},\ and\ \citenamefont
		{Segev}}]{Schwartz2008}%
	\BibitemOpen
	\bibfield  {author} {\bibinfo {author} {\bibfnamefont {T.}~\bibnamefont
			{Schwartz}}, \bibinfo {author} {\bibfnamefont {S.}~\bibnamefont {Fishman}}, \
		and\ \bibinfo {author} {\bibfnamefont {M.}~\bibnamefont {Segev}},\ }\bibfield
	{title} {\enquote {\bibinfo {title} {Localisation of light in disordered
				lattices},}\ }\href {\doibase 10.1049/el:20083646} {\bibfield  {journal}
		{\bibinfo  {journal} {Electron. Lett.}\ }\textbf {\bibinfo {volume} {44}},\
		\bibinfo {pages} {165} (\bibinfo {year} {2008})}\BibitemShut {NoStop}%
	\bibitem [{\citenamefont {Lahini}\ \emph {et~al.}(2008)\citenamefont {Lahini},
		\citenamefont {Avidan}, \citenamefont {Pozzi}, \citenamefont {Sorel},
		\citenamefont {Morandotti}, \citenamefont {Christodoulides},\ and\
		\citenamefont {Silberberg}}]{Lahini2008}%
	\BibitemOpen
	\bibfield  {author} {\bibinfo {author} {\bibfnamefont {Y.}~\bibnamefont
			{Lahini}}, \bibinfo {author} {\bibfnamefont {A.}~\bibnamefont {Avidan}},
		\bibinfo {author} {\bibfnamefont {F.}~\bibnamefont {Pozzi}}, \bibinfo
		{author} {\bibfnamefont {M.}~\bibnamefont {Sorel}}, \bibinfo {author}
		{\bibfnamefont {R.}~\bibnamefont {Morandotti}}, \bibinfo {author}
		{\bibfnamefont {D.~N.}\ \bibnamefont {Christodoulides}}, \ and\ \bibinfo
		{author} {\bibfnamefont {Y.}~\bibnamefont {Silberberg}},\ }\bibfield  {title}
	{\enquote {\bibinfo {title} {{A}nderson localization and nonlinearity in
				one-dimensional disordered photonic lattices},}\ }\href {\doibase
		10.1103/PhysRevLett.100.013906} {\bibfield  {journal} {\bibinfo  {journal}
			{Phys. Rev. Lett.}\ }\textbf {\bibinfo {volume} {100}},\ \bibinfo {pages}
		{013906} (\bibinfo {year} {2008})},\ \Eprint {http://arxiv.org/abs/0704.3788}
	{arXiv:0704.3788 [cond-mat.other]} \BibitemShut {NoStop}%
	\bibitem [{\citenamefont {Thompson}\ \emph {et~al.}(2010)\citenamefont
		{Thompson}, \citenamefont {Vemuri},\ and\ \citenamefont
		{Agarwal}}]{Thompson2010}%
	\BibitemOpen
	\bibfield  {author} {\bibinfo {author} {\bibfnamefont {C.}~\bibnamefont
			{Thompson}}, \bibinfo {author} {\bibfnamefont {G.}~\bibnamefont {Vemuri}}, \
		and\ \bibinfo {author} {\bibfnamefont {G.~S.}\ \bibnamefont {Agarwal}},\
	}\bibfield  {title} {\enquote {\bibinfo {title} {{A}nderson localization with
				second quantized fields in a coupled array of waveguides},}\ }\href {\doibase
		10.1103/PhysRevA.82.053805} {\bibfield  {journal} {\bibinfo  {journal} {Phys.
				Rev. A}\ }\textbf {\bibinfo {volume} {82}},\ \bibinfo {pages} {053805}
		(\bibinfo {year} {2010})}\BibitemShut {NoStop}%
	\bibitem [{\citenamefont {Longhi}(2011)}]{Longhi2011a}%
	\BibitemOpen
	\bibfield  {author} {\bibinfo {author} {\bibfnamefont {S.}~\bibnamefont
			{Longhi}},\ }\bibfield  {title} {\enquote {\bibinfo {title}
			{{J}aynes-{C}ummings photonic superlattices},}\ }\href {\doibase
		10.1364/OL.36.003407} {\bibfield  {journal} {\bibinfo  {journal} {Opt.
				Lett.}\ }\textbf {\bibinfo {volume} {36}},\ \bibinfo {pages} {3407} (\bibinfo
		{year} {2011})},\ \Eprint {http://arxiv.org/abs/1111.3457} {arXiv:1111.3457
		[physics.atom-ph]} \BibitemShut {NoStop}%
	\bibitem [{\citenamefont {Crespi}\ \emph {et~al.}(2012)\citenamefont {Crespi},
		\citenamefont {Longhi},\ and\ \citenamefont {Osellame}}]{Crespi2012}%
	\BibitemOpen
	\bibfield  {author} {\bibinfo {author} {\bibfnamefont {A.}~\bibnamefont
			{Crespi}}, \bibinfo {author} {\bibfnamefont {S.}~\bibnamefont {Longhi}}, \
		and\ \bibinfo {author} {\bibfnamefont {R.}~\bibnamefont {Osellame}},\
	}\bibfield  {title} {\enquote {\bibinfo {title} {Photonic realization of the
				quantum {R}abi model},}\ }\href {\doibase 10.1103/PhysRevLett.108.163601}
	{\bibfield  {journal} {\bibinfo  {journal} {Phys. Rev. Lett.}\ }\textbf
		{\bibinfo {volume} {108}},\ \bibinfo {pages} {163601} (\bibinfo {year}
		{2012})},\ \Eprint {http://arxiv.org/abs/1111.6424} {arXiv:1111.6424
		[quant-ph]} \BibitemShut {NoStop}%
	\bibitem [{\citenamefont {Rodr\'iguez-Lara}(2014)}]{RodriguezLara2014}%
	\BibitemOpen
	\bibfield  {author} {\bibinfo {author} {\bibfnamefont {B.~M.}\ \bibnamefont
			{Rodr\'iguez-Lara}},\ }\bibfield  {title} {\enquote {\bibinfo {title}
			{Intensity-dependent quantum {R}abi model: spectrum, supersymmetric partner,
				and optical simulation},}\ }\href {\doibase 10.1364/JOSAB.31.001719}
	{\bibfield  {journal} {\bibinfo  {journal} {J. Opt. Soc. Am. B}\ }\textbf
		{\bibinfo {volume} {31}},\ \bibinfo {pages} {1719} (\bibinfo {year}
		{2014})},\ \Eprint {http://arxiv.org/abs/1401.7376} {arXiv:1401.7376
		[quant-ph]} \BibitemShut {NoStop}%
	\bibitem [{\citenamefont {Snyder}(1972)}]{Snyder1972}%
	\BibitemOpen
	\bibfield  {author} {\bibinfo {author} {\bibfnamefont {A.~W.}\ \bibnamefont
			{Snyder}},\ }\bibfield  {title} {\enquote {\bibinfo {title} {Coupled-mode
				theory for optical fibers},}\ }\href {\doibase 10.1364/JOSA.62.001267}
	{\bibfield  {journal} {\bibinfo  {journal} {J. Opt. Soc. Am.}\ }\textbf
		{\bibinfo {volume} {62}},\ \bibinfo {pages} {1267} (\bibinfo {year}
		{1972})}\BibitemShut {NoStop}%
	\bibitem [{\citenamefont {McIntyre}\ and\ \citenamefont
		{Snyder}(1973)}]{McIntyre1973}%
	\BibitemOpen
	\bibfield  {author} {\bibinfo {author} {\bibfnamefont {P.~D.}\ \bibnamefont
			{McIntyre}}\ and\ \bibinfo {author} {\bibfnamefont {A.~W.}\ \bibnamefont
			{Snyder}},\ }\bibfield  {title} {\enquote {\bibinfo {title} {Power transfer
				between optical fibers},}\ }\href {\doibase 10.1364/josa.63.001518}
	{\bibfield  {journal} {\bibinfo  {journal} {J. Opt. Soc. Am.}\ }\textbf
		{\bibinfo {volume} {63}},\ \bibinfo {pages} {1518} (\bibinfo {year}
		{1973})}\BibitemShut {NoStop}%
	\bibitem [{\citenamefont {Huang}(1994)}]{Huang1994}%
	\BibitemOpen
	\bibfield  {author} {\bibinfo {author} {\bibfnamefont {W.-P.}\ \bibnamefont
			{Huang}},\ }\bibfield  {title} {\enquote {\bibinfo {title} {Coupled-mode
				theory for optical waveguides: an overview},}\ }\href {\doibase
		10.1364/JOSAA.11.000963} {\bibfield  {journal} {\bibinfo  {journal} {J. Opt.
				Soc. Am. A}\ }\textbf {\bibinfo {volume} {11}},\ \bibinfo {pages} {963}
		(\bibinfo {year} {1994})}\BibitemShut {NoStop}%
	\bibitem [{\citenamefont {Davis}\ \emph {et~al.}(1996)\citenamefont {Davis},
		\citenamefont {Miura}, \citenamefont {Sugimoto},\ and\ \citenamefont
		{Hirao}}]{Davis1996}%
	\BibitemOpen
	\bibfield  {author} {\bibinfo {author} {\bibfnamefont {K.~M.}\ \bibnamefont
			{Davis}}, \bibinfo {author} {\bibfnamefont {K.}~\bibnamefont {Miura}},
		\bibinfo {author} {\bibfnamefont {N.}~\bibnamefont {Sugimoto}}, \ and\
		\bibinfo {author} {\bibfnamefont {K.}~\bibnamefont {Hirao}},\ }\bibfield
	{title} {\enquote {\bibinfo {title} {Writing waveguides in glass with a
				femtosecond laser},}\ }\href {\doibase 10.1364/OL.21.001729} {\bibfield
		{journal} {\bibinfo  {journal} {Opt. Lett.}\ }\textbf {\bibinfo {volume}
			{21}},\ \bibinfo {pages} {1729} (\bibinfo {year} {1996})}\BibitemShut
	{NoStop}%
	\bibitem [{\citenamefont {Bl\"omer}\ \emph {et~al.}(2006)\citenamefont
		{Bl\"omer}, \citenamefont {Szameit}, \citenamefont {Dreisow}, \citenamefont
		{Schreiber}, \citenamefont {Nolte},\ and\ \citenamefont
		{T\"unnermann}}]{Blomer2006}%
	\BibitemOpen
	\bibfield  {author} {\bibinfo {author} {\bibfnamefont {D.}~\bibnamefont
			{Bl\"omer}}, \bibinfo {author} {\bibfnamefont {A.}~\bibnamefont {Szameit}},
		\bibinfo {author} {\bibfnamefont {F.}~\bibnamefont {Dreisow}}, \bibinfo
		{author} {\bibfnamefont {T.}~\bibnamefont {Schreiber}}, \bibinfo {author}
		{\bibfnamefont {S.}~\bibnamefont {Nolte}}, \ and\ \bibinfo {author}
		{\bibfnamefont {A.}~\bibnamefont {T\"unnermann}},\ }\bibfield  {title}
	{\enquote {\bibinfo {title} {Nonlinear refractive index of fs-laser-written
				waveguides in fused silica},}\ }\href {\doibase 10.1364/OE.14.002151}
	{\bibfield  {journal} {\bibinfo  {journal} {Opt. Express}\ }\textbf {\bibinfo
			{volume} {14}},\ \bibinfo {pages} {2151} (\bibinfo {year}
		{2006})}\BibitemShut {NoStop}%
	\bibitem [{\citenamefont {Szameit}\ and\ \citenamefont
		{Nolte}(2010)}]{Szameit2010}%
	\BibitemOpen
	\bibfield  {author} {\bibinfo {author} {\bibfnamefont {A.}~\bibnamefont
			{Szameit}}\ and\ \bibinfo {author} {\bibfnamefont {S.}~\bibnamefont
			{Nolte}},\ }\bibfield  {title} {\enquote {\bibinfo {title} {Discrete optics
				in femtosecond-laser-written photonic structures},}\ }\href {\doibase
		10.1088/0953-4075/43/16/163001} {\bibfield  {journal} {\bibinfo  {journal}
			{J. Phys. B: At. Mol. Opt. Phys.}\ }\textbf {\bibinfo {volume} {43}},\
		\bibinfo {pages} {163001} (\bibinfo {year} {2010})}\BibitemShut {NoStop}%
	\bibitem [{\citenamefont {Leggett}\ \emph {et~al.}(1987)\citenamefont
		{Leggett}, \citenamefont {Chakravarty}, \citenamefont {Dorsey}, \citenamefont
		{Fisher}, \citenamefont {Garg},\ and\ \citenamefont {Zwerger}}]{Leggett1987}%
	\BibitemOpen
	\bibfield  {author} {\bibinfo {author} {\bibfnamefont {A.~J.}\ \bibnamefont
			{Leggett}}, \bibinfo {author} {\bibfnamefont {S.}~\bibnamefont
			{Chakravarty}}, \bibinfo {author} {\bibfnamefont {A.~T.}\ \bibnamefont
			{Dorsey}}, \bibinfo {author} {\bibfnamefont {M.~P.~A.}\ \bibnamefont
			{Fisher}}, \bibinfo {author} {\bibfnamefont {A.}~\bibnamefont {Garg}}, \ and\
		\bibinfo {author} {\bibfnamefont {W.}~\bibnamefont {Zwerger}},\ }\bibfield
	{title} {\enquote {\bibinfo {title} {Dynamics of the dissipative two-state
				system},}\ }\href {\doibase 10.1103/RevModPhys.59.1} {\bibfield  {journal}
		{\bibinfo  {journal} {Rev. Mod. Phys.}\ }\textbf {\bibinfo {volume} {59}},\
		\bibinfo {pages} {1} (\bibinfo {year} {1987})}\BibitemShut {NoStop}%
	\bibitem [{\citenamefont {Weiss}(2009)}]{Weiss2009}%
	\BibitemOpen
	\bibfield  {author} {\bibinfo {author} {\bibfnamefont {U.}~\bibnamefont
			{Weiss}},\ }\href {\doibase 10.1142/8334} {\emph {\bibinfo {title} {Quantum
				dissipative systems}}},\ \bibinfo {edition} {4th}\ ed.\ (\bibinfo
	{publisher} {World Scientific},\ \bibinfo {address} {Singapore},\ \bibinfo
	{year} {2009})\BibitemShut {NoStop}%
	\bibitem [{\citenamefont {Shiokawa}\ and\ \citenamefont
		{Hu}(2004)}]{Shiokawa2004}%
	\BibitemOpen
	\bibfield  {author} {\bibinfo {author} {\bibfnamefont {K.}~\bibnamefont
			{Shiokawa}}\ and\ \bibinfo {author} {\bibfnamefont {B.~L.}\ \bibnamefont
			{Hu}},\ }\bibfield  {title} {\enquote {\bibinfo {title} {Qubit decoherence
				and non-{M}arkovian dynamics at low temperatures via an effective spin-boson
				model},}\ }\href {\doibase 10.1103/PhysRevA.70.062106} {\bibfield  {journal}
		{\bibinfo  {journal} {Phys. Rev. A}\ }\textbf {\bibinfo {volume} {70}},\
		\bibinfo {pages} {062106} (\bibinfo {year} {2004})},\ \Eprint
	{http://arxiv.org/abs/quant-ph/0405147} {arXiv:quant-ph/0405147} \BibitemShut
	{NoStop}%
	\bibitem [{\citenamefont {Guarnieri}\ \emph {et~al.}(2016)\citenamefont
		{Guarnieri}, \citenamefont {Uchiyama},\ and\ \citenamefont
		{Vacchini}}]{Guarnieri2016}%
	\BibitemOpen
	\bibfield  {author} {\bibinfo {author} {\bibfnamefont {G.}~\bibnamefont
			{Guarnieri}}, \bibinfo {author} {\bibfnamefont {C.}~\bibnamefont {Uchiyama}},
		\ and\ \bibinfo {author} {\bibfnamefont {B.}~\bibnamefont {Vacchini}},\
	}\bibfield  {title} {\enquote {\bibinfo {title} {Energy backflow and
				non-{M}arkovian dynamics},}\ }\href {\doibase 10.1103/PhysRevA.93.012118}
	{\bibfield  {journal} {\bibinfo  {journal} {Phys. Rev. A}\ }\textbf {\bibinfo
			{volume} {93}},\ \bibinfo {pages} {012118} (\bibinfo {year} {2016})},\
	\Eprint {http://arxiv.org/abs/1510.02333} {arXiv:1510.02333 [quant-ph]}
	\BibitemShut {NoStop}%
	\bibitem [{\citenamefont {Vojta}\ \emph {et~al.}(2005)\citenamefont {Vojta},
		\citenamefont {Tong},\ and\ \citenamefont {Bulla}}]{Vojta2005}%
	\BibitemOpen
	\bibfield  {author} {\bibinfo {author} {\bibfnamefont {M.}~\bibnamefont
			{Vojta}}, \bibinfo {author} {\bibfnamefont {N.-H.}\ \bibnamefont {Tong}}, \
		and\ \bibinfo {author} {\bibfnamefont {R.}~\bibnamefont {Bulla}},\ }\bibfield
	{title} {\enquote {\bibinfo {title} {Quantum phase transitions in the
				sub-ohmic spin-boson model: Failure of the quantum-classical mapping},}\
	}\href {\doibase 10.1103/PhysRevLett.94.070604} {\bibfield  {journal}
		{\bibinfo  {journal} {Phys. Rev. Lett.}\ }\textbf {\bibinfo {volume} {94}},\
		\bibinfo {pages} {070604} (\bibinfo {year} {2005})},\ \Eprint
	{http://arxiv.org/abs/cond-mat/0410132} {arXiv:cond-mat/0410132} \BibitemShut
	{NoStop}%
	\bibitem [{\citenamefont {Chin}\ \emph {et~al.}(2010)\citenamefont {Chin},
		\citenamefont {Rivas}, \citenamefont {Huelga},\ and\ \citenamefont
		{Plenio}}]{Chin2010}%
	\BibitemOpen
	\bibfield  {author} {\bibinfo {author} {\bibfnamefont {A.~W.}\ \bibnamefont
			{Chin}}, \bibinfo {author} {\bibfnamefont {A.}~\bibnamefont {Rivas}},
		\bibinfo {author} {\bibfnamefont {S.~F.}\ \bibnamefont {Huelga}}, \ and\
		\bibinfo {author} {\bibfnamefont {M.~B.}\ \bibnamefont {Plenio}},\ }\bibfield
	{title} {\enquote {\bibinfo {title} {Exact mapping between system-reservoir
				quantum models and semi-infinite discrete chains using orthogonal
				polynomials},}\ }\href {\doibase 10.1063/1.3490188} {\bibfield  {journal}
		{\bibinfo  {journal} {J. Math. Phys.}\ }\textbf {\bibinfo {volume} {51}},\
		\bibinfo {pages} {092109} (\bibinfo {year} {2010})},\ \Eprint
	{http://arxiv.org/abs/1006.4507} {arXiv:1006.4507 [quant-ph]} \BibitemShut
	{NoStop}%
	\bibitem [{\citenamefont {Prior}\ \emph {et~al.}(2010)\citenamefont {Prior},
		\citenamefont {Chin}, \citenamefont {Huelga},\ and\ \citenamefont
		{Plenio}}]{Prior2010}%
	\BibitemOpen
	\bibfield  {author} {\bibinfo {author} {\bibfnamefont {J.}~\bibnamefont
			{Prior}}, \bibinfo {author} {\bibfnamefont {A.~W.}\ \bibnamefont {Chin}},
		\bibinfo {author} {\bibfnamefont {S.~F.}\ \bibnamefont {Huelga}}, \ and\
		\bibinfo {author} {\bibfnamefont {M.~B.}\ \bibnamefont {Plenio}},\ }\bibfield
	{title} {\enquote {\bibinfo {title} {Efficient simulation of strong
				system-environment interactions},}\ }\href {\doibase
		10.1103/PhysRevLett.105.050404} {\bibfield  {journal} {\bibinfo  {journal}
			{Phys. Rev. Lett.}\ }\textbf {\bibinfo {volume} {105}},\ \bibinfo {pages}
		{050404} (\bibinfo {year} {2010})},\ \Eprint {http://arxiv.org/abs/1003.5503}
	{arXiv:1003.5503 [quant-ph]} \BibitemShut {NoStop}%
	\bibitem [{\citenamefont {Woods}\ \emph {et~al.}(2014)\citenamefont {Woods},
		\citenamefont {Groux}, \citenamefont {Chin}, \citenamefont {Huelga},\ and\
		\citenamefont {Plenio}}]{Woods2014}%
	\BibitemOpen
	\bibfield  {author} {\bibinfo {author} {\bibfnamefont {M.~P.}\ \bibnamefont
			{Woods}}, \bibinfo {author} {\bibfnamefont {R.}~\bibnamefont {Groux}},
		\bibinfo {author} {\bibfnamefont {A.~W.}\ \bibnamefont {Chin}}, \bibinfo
		{author} {\bibfnamefont {S.~F.}\ \bibnamefont {Huelga}}, \ and\ \bibinfo
		{author} {\bibfnamefont {M.~B.}\ \bibnamefont {Plenio}},\ }\bibfield  {title}
	{\enquote {\bibinfo {title} {Mappings of open quantum systems onto chain
				representations and {M}arkovian embeddings},}\ }\href {\doibase
		10.1063/1.4866769} {\bibfield  {journal} {\bibinfo  {journal} {J. Math.
				Phys.}\ }\textbf {\bibinfo {volume} {55}},\ \bibinfo {pages} {032101}
		(\bibinfo {year} {2014})},\ \Eprint {http://arxiv.org/abs/1111.5262}
	{arXiv:1111.5262 [quant-ph]} \BibitemShut {NoStop}%
	\bibitem [{\citenamefont {Meade}\ \emph {et~al.}(1995)\citenamefont {Meade},
		\citenamefont {Winn},\ and\ \citenamefont {Joannopoulos}}]{Meade1995}%
	\BibitemOpen
	\bibfield  {author} {\bibinfo {author} {\bibfnamefont {R.~D.}\ \bibnamefont
			{Meade}}, \bibinfo {author} {\bibfnamefont {J.~N.}\ \bibnamefont {Winn}}, \
		and\ \bibinfo {author} {\bibfnamefont {J.~D.}\ \bibnamefont {Joannopoulos}},\
	}\href@noop {} {\emph {\bibinfo {title} {Photonic crystals: Molding the flow
				of light}}},\ \bibinfo {edition} {2nd}\ ed.\ (\bibinfo  {publisher}
	{Princeton University Press},\ \bibinfo {address} {U.S.A.},\ \bibinfo {year}
	{1995})\BibitemShut {NoStop}%
	\bibitem [{\citenamefont {Garanovich}\ \emph {et~al.}(2012)\citenamefont
		{Garanovich}, \citenamefont {Longhi}, \citenamefont {Sukhorukov},\ and\
		\citenamefont {Kivshar}}]{Garanovich2012}%
	\BibitemOpen
	\bibfield  {author} {\bibinfo {author} {\bibfnamefont {I.~L.}\ \bibnamefont
			{Garanovich}}, \bibinfo {author} {\bibfnamefont {S.}~\bibnamefont {Longhi}},
		\bibinfo {author} {\bibfnamefont {A.~A.}\ \bibnamefont {Sukhorukov}}, \ and\
		\bibinfo {author} {\bibfnamefont {Y.~S.}\ \bibnamefont {Kivshar}},\
	}\bibfield  {title} {\enquote {\bibinfo {title} {Light propagation and
				localization in modulated photonic lattices and waveguides},}\ }\href
	{\doibase 10.1016/j.physrep.2012.03.005} {\bibfield  {journal} {\bibinfo
			{journal} {Phys. Rep.}\ }\textbf {\bibinfo {volume} {518}},\ \bibinfo {pages}
		{1} (\bibinfo {year} {2012})},\ \Eprint {http://arxiv.org/abs/1107.2992}
	{arXiv:1107.2992 [physics.optics]} \BibitemShut {NoStop}%
	\bibitem [{\citenamefont {Longhi}(2007{\natexlab{a}})}]{Longhi2007}%
	\BibitemOpen
	\bibfield  {author} {\bibinfo {author} {\bibfnamefont {S.}~\bibnamefont
			{Longhi}},\ }\bibfield  {title} {\enquote {\bibinfo {title} {Bound states in
				the continuum in a single-level {F}ano-{A}nderson model},}\ }\href {\doibase
		10.1140/epjb/e2007-00143-2} {\bibfield  {journal} {\bibinfo  {journal} {Eur.
				Phys. J.}\ }\textbf {\bibinfo {volume} {57}},\ \bibinfo {pages} {45}
		(\bibinfo {year} {2007}{\natexlab{a}})}\BibitemShut {NoStop}%
	\bibitem [{\citenamefont {Plotnik}\ \emph {et~al.}(2011)\citenamefont
		{Plotnik}, \citenamefont {Peleg}, \citenamefont {Dreisow}, \citenamefont
		{Heinrich}, \citenamefont {Nolte}, \citenamefont {Szameit},\ and\
		\citenamefont {Segev}}]{Plotnik2011}%
	\BibitemOpen
	\bibfield  {author} {\bibinfo {author} {\bibfnamefont {Y.}~\bibnamefont
			{Plotnik}}, \bibinfo {author} {\bibfnamefont {O.}~\bibnamefont {Peleg}},
		\bibinfo {author} {\bibfnamefont {F.}~\bibnamefont {Dreisow}}, \bibinfo
		{author} {\bibfnamefont {M.}~\bibnamefont {Heinrich}}, \bibinfo {author}
		{\bibfnamefont {S.}~\bibnamefont {Nolte}}, \bibinfo {author} {\bibfnamefont
			{A.}~\bibnamefont {Szameit}}, \ and\ \bibinfo {author} {\bibfnamefont
			{M.}~\bibnamefont {Segev}},\ }\bibfield  {title} {\enquote {\bibinfo {title}
			{Experimental observation of optical bound states in the continuum},}\ }\href
	{\doibase 10.1103/PhysRevLett.107.183901} {\bibfield  {journal} {\bibinfo
			{journal} {Phys. Rev. Lett.}\ }\textbf {\bibinfo {volume} {107}},\ \bibinfo
		{pages} {183901} (\bibinfo {year} {2011})}\BibitemShut {NoStop}%
	\bibitem [{\citenamefont {Dreisow}\ \emph {et~al.}(2008)\citenamefont
		{Dreisow}, \citenamefont {Szameit}, \citenamefont {Heinrich}, \citenamefont
		{Pertsch}, \citenamefont {Nolte}, \citenamefont {T\"unnermann},\ and\
		\citenamefont {Longhi}}]{Dreisow2008}%
	\BibitemOpen
	\bibfield  {author} {\bibinfo {author} {\bibfnamefont {F.}~\bibnamefont
			{Dreisow}}, \bibinfo {author} {\bibfnamefont {A.}~\bibnamefont {Szameit}},
		\bibinfo {author} {\bibfnamefont {M.}~\bibnamefont {Heinrich}}, \bibinfo
		{author} {\bibfnamefont {T.}~\bibnamefont {Pertsch}}, \bibinfo {author}
		{\bibfnamefont {S.}~\bibnamefont {Nolte}}, \bibinfo {author} {\bibfnamefont
			{A.}~\bibnamefont {T\"unnermann}}, \ and\ \bibinfo {author} {\bibfnamefont
			{S.}~\bibnamefont {Longhi}},\ }\bibfield  {title} {\enquote {\bibinfo {title}
			{Decay control via discrete-to-continuum coupling modulation in an optical
				waveguide system},}\ }\href {\doibase 10.1103/PhysRevLett.101.143602}
	{\bibfield  {journal} {\bibinfo  {journal} {Phys. Rev. Lett.}\ }\textbf
		{\bibinfo {volume} {101}},\ \bibinfo {pages} {143602} (\bibinfo {year}
		{2008})}\BibitemShut {NoStop}%
	\bibitem [{\citenamefont {Longhi}(2009{\natexlab{b}})}]{Longhi2009a}%
	\BibitemOpen
	\bibfield  {author} {\bibinfo {author} {\bibfnamefont {S.}~\bibnamefont
			{Longhi}},\ }\bibfield  {title} {\enquote {\bibinfo {title} {Optical analogue
				of coherent population trapping via a continuum in optical waveguide
				arrays},}\ }\href {\doibase 10.1080/09500340802187373} {\bibfield  {journal}
		{\bibinfo  {journal} {J. Mod. Optic.}\ }\textbf {\bibinfo {volume} {56}},\
		\bibinfo {pages} {729} (\bibinfo {year} {2009}{\natexlab{b}})}\BibitemShut
	{NoStop}%
	\bibitem [{\citenamefont {Longhi}(2006)}]{Longhi2006a}%
	\BibitemOpen
	\bibfield  {author} {\bibinfo {author} {\bibfnamefont {S.}~\bibnamefont
			{Longhi}},\ }\bibfield  {title} {\enquote {\bibinfo {title} {Nonexponential
				decay via tunneling in tight-binding lattices and the optical {Z}eno
				effect},}\ }\href {\doibase 10.1002/prop.201200077} {\bibfield  {journal}
		{\bibinfo  {journal} {Phys. Rev. Lett.}\ }\textbf {\bibinfo {volume} {97}},\
		\bibinfo {pages} {110402} (\bibinfo {year} {2006})}\BibitemShut {NoStop}%
	\bibitem [{\citenamefont {Longhi}(2007{\natexlab{b}})}]{Longhi2007a}%
	\BibitemOpen
	\bibfield  {author} {\bibinfo {author} {\bibfnamefont {S.}~\bibnamefont
			{Longhi}},\ }\bibfield  {title} {\enquote {\bibinfo {title} {Control of
				photon tunneling in optical waveguides},}\ }\href {\doibase
		10.1364/OL.32.000557} {\bibfield  {journal} {\bibinfo  {journal} {Opt.
				Lett.}\ }\textbf {\bibinfo {volume} {32}},\ \bibinfo {pages} {557} (\bibinfo
		{year} {2007}{\natexlab{b}})}\BibitemShut {NoStop}%
	\bibitem [{\citenamefont {Biagioni}\ \emph {et~al.}(2008)\citenamefont
		{Biagioni}, \citenamefont {{Della Valle}}, \citenamefont {Ornigotti},
		\citenamefont {Finazzi}, \citenamefont {Duo}, \citenamefont {Laporta},\ and\
		\citenamefont {Longhi}}]{Biagioni2008}%
	\BibitemOpen
	\bibfield  {author} {\bibinfo {author} {\bibfnamefont {P.}~\bibnamefont
			{Biagioni}}, \bibinfo {author} {\bibfnamefont {G.}~\bibnamefont {{Della
					Valle}}}, \bibinfo {author} {\bibfnamefont {M.}~\bibnamefont {Ornigotti}},
		\bibinfo {author} {\bibfnamefont {M.}~\bibnamefont {Finazzi}}, \bibinfo
		{author} {\bibfnamefont {L.}~\bibnamefont {Duo}}, \bibinfo {author}
		{\bibfnamefont {P.}~\bibnamefont {Laporta}}, \ and\ \bibinfo {author}
		{\bibfnamefont {S.}~\bibnamefont {Longhi}},\ }\bibfield  {title} {\enquote
		{\bibinfo {title} {Experimental demonstration of the optical {Z}eno effect by
				scanning tunneling optical microscopy},}\ }\href {\doibase
		10.1364/OE.16.003762} {\bibfield  {journal} {\bibinfo  {journal} {Opt.
				Express}\ }\textbf {\bibinfo {volume} {16}},\ \bibinfo {pages} {3762}
		(\bibinfo {year} {2008})}\BibitemShut {NoStop}%
	\bibitem [{\citenamefont {Kockum}\ \emph {et~al.}(2018)\citenamefont {Kockum},
		\citenamefont {Johansson},\ and\ \citenamefont {Nori}}]{Kockum2018}%
	\BibitemOpen
	\bibfield  {author} {\bibinfo {author} {\bibfnamefont {A.~F.}\ \bibnamefont
			{Kockum}}, \bibinfo {author} {\bibfnamefont {G.}~\bibnamefont {Johansson}}, \
		and\ \bibinfo {author} {\bibfnamefont {F.}~\bibnamefont {Nori}},\ }\bibfield
	{title} {\enquote {\bibinfo {title} {Decoherence-free interaction between
				giant atoms in waveguide quantum electrodynamics},}\ }\href {\doibase
		10.1103/PhysRevLett.120.140404} {\bibfield  {journal} {\bibinfo  {journal}
			{Phys. Rev. Lett.}\ }\textbf {\bibinfo {volume} {120}},\ \bibinfo {pages}
		{140404} (\bibinfo {year} {2018})},\ \Eprint
	{http://arxiv.org/abs/1711.08863} {arXiv:1711.08863 [quant-ph]} \BibitemShut
	{NoStop}%
	\bibitem [{\citenamefont {Longhi}(2020)}]{Longhi2020}%
	\BibitemOpen
	\bibfield  {author} {\bibinfo {author} {\bibfnamefont {S.}~\bibnamefont
			{Longhi}},\ }\bibfield  {title} {\enquote {\bibinfo {title} {Photonic
				simulation of giant atom decay},}\ }\href {\doibase 10.1364/OL.393578}
	{\bibfield  {journal} {\bibinfo  {journal} {Opt. Lett.}\ }\textbf {\bibinfo
			{volume} {45}},\ \bibinfo {pages} {3017} (\bibinfo {year}
		{2020})}\BibitemShut {NoStop}%
	\bibitem [{\citenamefont {Gross}\ and\ \citenamefont
		{Withford}(2015)}]{Gross2015}%
	\BibitemOpen
	\bibfield  {author} {\bibinfo {author} {\bibfnamefont {S.}~\bibnamefont
			{Gross}}\ and\ \bibinfo {author} {\bibfnamefont {M.~J.}\ \bibnamefont
			{Withford}},\ }\bibfield  {title} {\enquote {\bibinfo {title}
			{Ultrafast-laser-inscribed 3{D} integrated photonics: challenges and emerging
				applications},}\ }\href {\doibase https://doi.org/10.1515/nanoph-2015-0020}
	{\bibfield  {journal} {\bibinfo  {journal} {Nanophotonics}\ }\textbf
		{\bibinfo {volume} {4}},\ \bibinfo {pages} {332} (\bibinfo {year}
		{2015})}\BibitemShut {NoStop}%
	\bibitem [{\citenamefont {H\"{o}rmander}(1990)}]{Hormander}%
	\BibitemOpen
	\bibfield  {author} {\bibinfo {author} {\bibfnamefont {L.}~\bibnamefont
			{H\"{o}rmander}},\ }\href {\doibase 10.1007/978-3-642-61497-2} {\emph
		{\bibinfo {title} {The Analysis of Linear Partial Differential Operators. I,
				Distribution Theory and {F}ourier Analysis}}},\ \bibinfo {edition} {2nd}\
	ed.,\ Grundlehren Der Mathematischen Wissenschaften\ (\bibinfo  {publisher}
	{Springer},\ \bibinfo {address} {Germany},\ \bibinfo {year}
	{1990})\BibitemShut {NoStop}%
	\bibitem [{\citenamefont {Visuri}\ \emph {et~al.}(2018)\citenamefont {Visuri},
		\citenamefont {Berthod},\ and\ \citenamefont {Giamarchi}}]{Visuri2018}%
	\BibitemOpen
	\bibfield  {author} {\bibinfo {author} {\bibfnamefont {A.-M.}\ \bibnamefont
			{Visuri}}, \bibinfo {author} {\bibfnamefont {C.}~\bibnamefont {Berthod}}, \
		and\ \bibinfo {author} {\bibfnamefont {T.}~\bibnamefont {Giamarchi}},\
	}\bibfield  {title} {\enquote {\bibinfo {title} {Impurity coupled to a
				lattice with disorder},}\ }\href {\doibase 10.1103/PhysRevA.98.053607}
	{\bibfield  {journal} {\bibinfo  {journal} {Phys. Rev. A}\ }\textbf {\bibinfo
			{volume} {98}},\ \bibinfo {pages} {053607} (\bibinfo {year} {2018})},\
	\Eprint {http://arxiv.org/abs/1807.07744} {arXiv:1807.07744
		[cond-mat.quant-gas]} \BibitemShut {NoStop}%
	\bibitem [{\citenamefont {{De Raedt}}\ \emph {et~al.}(1989)\citenamefont {{De
				Raedt}}, \citenamefont {Lagendijk},\ and\ \citenamefont {{de
				Vries}}}]{DeRaedt1989}%
	\BibitemOpen
	\bibfield  {author} {\bibinfo {author} {\bibfnamefont {H.}~\bibnamefont {{De
					Raedt}}}, \bibinfo {author} {\bibfnamefont {A.}~\bibnamefont {Lagendijk}}, \
		and\ \bibinfo {author} {\bibfnamefont {P.}~\bibnamefont {{de Vries}}},\
	}\bibfield  {title} {\enquote {\bibinfo {title} {Transverse localization of
				light},}\ }\href {\doibase 10.1103/PhysRevLett.62.47} {\bibfield  {journal}
		{\bibinfo  {journal} {Phys. Rev. Lett.}\ }\textbf {\bibinfo {volume} {62}},\
		\bibinfo {pages} {47} (\bibinfo {year} {1989})}\BibitemShut {NoStop}%
	\bibitem [{\citenamefont {Schwartz}\ \emph {et~al.}(2007)\citenamefont
		{Schwartz}, \citenamefont {Bartal}, \citenamefont {Fishman},\ and\
		\citenamefont {Segev}}]{Schwartz2007}%
	\BibitemOpen
	\bibfield  {author} {\bibinfo {author} {\bibfnamefont {T.}~\bibnamefont
			{Schwartz}}, \bibinfo {author} {\bibfnamefont {G.}~\bibnamefont {Bartal}},
		\bibinfo {author} {\bibfnamefont {S.}~\bibnamefont {Fishman}}, \ and\
		\bibinfo {author} {\bibfnamefont {M.}~\bibnamefont {Segev}},\ }\bibfield
	{title} {\enquote {\bibinfo {title} {Transport and {A}nderson localization in
				disordered two-dimensional photonic lattices},}\ }\href {\doibase
		10.1038/nature05623} {\bibfield  {journal} {\bibinfo  {journal} {Nature}\
		}\textbf {\bibinfo {volume} {446}},\ \bibinfo {pages} {52} (\bibinfo {year}
		{2007})}\BibitemShut {NoStop}%
	\bibitem [{\citenamefont {Segev}\ \emph {et~al.}(2013)\citenamefont {Segev},
		\citenamefont {Silberberg},\ and\ \citenamefont
		{Christodoulides}}]{Segev2013}%
	\BibitemOpen
	\bibfield  {author} {\bibinfo {author} {\bibfnamefont {M.}~\bibnamefont
			{Segev}}, \bibinfo {author} {\bibfnamefont {Y.}~\bibnamefont {Silberberg}}, \
		and\ \bibinfo {author} {\bibfnamefont {D.~N.}\ \bibnamefont
			{Christodoulides}},\ }\bibfield  {title} {\enquote {\bibinfo {title}
			{{A}nderson localization of light},}\ }\href {\doibase
		10.1038/nphoton.2013.30} {\bibfield  {journal} {\bibinfo  {journal} {Nat.
				Photonics}\ }\textbf {\bibinfo {volume} {7}},\ \bibinfo {pages} {197}
		(\bibinfo {year} {2013})}\BibitemShut {NoStop}%
	\bibitem [{\citenamefont {{Jaramillo \'Avila}}\ \emph
		{et~al.}(2019)\citenamefont {{Jaramillo \'Avila}}, \citenamefont {Torres},
		\citenamefont {de~J.~Le\'on-Montiel},\ and\ \citenamefont
		{Rodr\'iguez-Lara}}]{Jaramillo2019}%
	\BibitemOpen
	\bibfield  {author} {\bibinfo {author} {\bibfnamefont {B.}~\bibnamefont
			{{Jaramillo \'Avila}}}, \bibinfo {author} {\bibfnamefont {J.~M.}\
			\bibnamefont {Torres}}, \bibinfo {author} {\bibfnamefont {R.}~\bibnamefont
			{de~J.~Le\'on-Montiel}}, \ and\ \bibinfo {author} {\bibfnamefont {B.~M.}\
			\bibnamefont {Rodr\'iguez-Lara}},\ }\bibfield  {title} {\enquote {\bibinfo
			{title} {Optimal crosstalk suppression in multicore fibers},}\ }\href
	{\doibase 10.1038/s41598-019-51854-x} {\bibfield  {journal} {\bibinfo
			{journal} {Sci. Rep.}\ }\textbf {\bibinfo {volume} {9}},\ \bibinfo {pages}
		{15737} (\bibinfo {year} {2019})},\ \Eprint {http://arxiv.org/abs/1905.09416}
	{arXiv:1905.09416 [physics.optics]} \BibitemShut {NoStop}%
	\bibitem [{\citenamefont {Alberucci}\ \emph {et~al.}(2020)\citenamefont
		{Alberucci}, \citenamefont {Alasgarzade}, \citenamefont {Chambonneau},
		\citenamefont {K\"ammer}, \citenamefont {Matth\"aus}, \citenamefont {Jisha},\
		and\ \citenamefont {Nolte}}]{Alberucci2020}%
	\BibitemOpen
	\bibfield  {author} {\bibinfo {author} {\bibfnamefont {A.}~\bibnamefont
			{Alberucci}}, \bibinfo {author} {\bibfnamefont {N.}~\bibnamefont
			{Alasgarzade}}, \bibinfo {author} {\bibfnamefont {M.}~\bibnamefont
			{Chambonneau}}, \bibinfo {author} {\bibfnamefont {M.~Blotheand~H.}\
			\bibnamefont {K\"ammer}}, \bibinfo {author} {\bibfnamefont {G.}~\bibnamefont
			{Matth\"aus}}, \bibinfo {author} {\bibfnamefont {C.~P.}\ \bibnamefont
			{Jisha}}, \ and\ \bibinfo {author} {\bibfnamefont {S.}~\bibnamefont
			{Nolte}},\ }\bibfield  {title} {\enquote {\bibinfo {title} {In-depth optical
				characterization of femtosecond-written waveguides in silicon},}\ }\href
	{\doibase 10.1103/PhysRevApplied.14.024078} {\bibfield  {journal} {\bibinfo
			{journal} {Phys. Rev. Appl.}\ }\textbf {\bibinfo {volume} {14}},\ \bibinfo
		{pages} {024078} (\bibinfo {year} {2020})}\BibitemShut {NoStop}%
\end{thebibliography}
%

\end{document}